\documentclass[12pt,a4paper,english,fleqn]{article}

\usepackage[sort&compress,round,comma,authoryear]{natbib}
\usepackage[T1]{fontenc}
\usepackage{ae,aecompl}

\usepackage[colorlinks=true,urlcolor=blue,citecolor=blue,linkcolor=red,bookmarks=true]{hyperref}
\usepackage{graphicx}
\usepackage{float}	
\usepackage{amsmath}	
\usepackage{amssymb}	
\newlength{\lp}
\usepackage[mathscr]{eucal}
\usepackage{amsfonts}
\makeatother
\usepackage{babel}
\makeatother
\begin{document}
\title{Modelling the SPARC galaxies using neo-MOND scaling relationships: the determination of distance scales and mass distributions purely from disk dynamical data. }
\author{D. F. Roscoe (The Open University; D.Roscoe@open.ac.uk)\\ \\ORCID: 0000-0003-3561-7425}
\date{}
\maketitle
\newpage
\begin{abstract}
The SPARC sample consists of 175 nearby galaxies with modern surface photometry at $3.6\,\mu m$ and high quality rotation curves. The sample has been constructed to span very wide ranges in surface brightness, luminosity, rotation velocity and Hubble type, thereby forming a representative sample of galaxies in the nearby Universe. To date, the SPARC sample is the largest collection of galaxies with both high-quality rotation curves and NIR surface photometry.
\\\\
Neo-MOND, used here to analyse the SPARC sample, recognizably conforms to the general pattern of the classical MOND algorithm, with the primary difference that, whereas classical MOND is purely empirical,  neo-MOND takes the form of a two-component scaling relation which arises as a special case solution of a cosmology motivated by the ideas of Leibniz and Mach (not discussed here).  The consequent main results can be broadly stated as follows: 
\begin{itemize}
	\item the neo-MOND two-component scaling relationship  provides for the derivation of a complete theory of the baryonic Tully-Fisher relation;
	\item the details of the derivation provide a means of setting absolute distance scales for disk galaxies independently of standard candles and the photometric method;
	\item subsequent determinations of whole-disk dynamical mass (computed directly from neo-MOND fits to SPARC rotation curves) track whole-disk photometric mass (estimated from SPARC surface photometry) across the whole SPARC sample in a statistically perfect way.
\end{itemize}
To summarize, if the input to the neo-MOND scaling relationships is whole-disk dynamical data, then the output is whole-disk mass data together with absolute distance scales.
\end{abstract}
\newpage{}


\section{Introduction:}\label{Intro}
The SPARC sample (\citet{McGaugh2015}) was made publically available in 2018 with the express intention of providing a tool for the study of mass distributions across the generality of spiral galaxies. It is the most comprehensive and accurate sample available for such purposes by a considerable margin, and its recent release into the public domain has made possible the analysis presented here.
\\\\
In practice, modern ideas about mass-modeling within galaxies fall into one of two categories: 
\begin{itemize}
	\item the general concensus is that some form or other of Dark Matter is essential if the observed dynamics within (generally) spiral galaxies is ever to be understood;
	\item resisting the general concensus is the very much minority view that a modification of the classical Newtonian theory is required - a view that is encapsulated within the MOND algorithm, introduced by Milgrom in the 1980s.
\end{itemize}
Milgrom, along with several other authors over the years, puzzled over the dual mysteries of the \lq{flat rotation curve}' phenomenon of disk galaxies and the baryonic Tully-Fisher relationship which related the asymptotic (flat) rotation velocity in such galaxies to their visible mass. His crucial insight in the early 1980's was the recognition that if the flip to flatness of rotation curves occurred on an \emph{acceleration} scale, rather than some distance scale which many had tried, then the baryonic Tully-Fisher relationship would follow as a natural conequence. This idea, MOND, (for Modified Newtonian Dynamics) proved to be surprisingly productive, as is evidenced by the blitz of work which followed \citet{Milgrom1983a, Milgrom1983b,Milgrom1983c,Milgrom1983d,Milgrom1984,Milgrom1988,Milgrom1989a,Milgrom1989b,Milgrom1989c,Milgrom1991,Milgrom1994a,Milgrom1994b,Milgrom1995,Milgrom1997a,Milgrom1997b,Milgrom1998,Milgrom1999}. Sanders \citet{Sanders1984, Sanders1986, Sanders1988, Sanders1989, Sanders1990, Sanders1994a, Sanders1994b, Sanders1996, Sanders1997, Sanders1998a, Sanders1998b, Sanders1999, Sanders2000, Sanders2001, Sanders2014},  McGaugh  \citet{McGaugh1995a,McGaugh1995b,McGaugh1996,McGaugh1998a,McGaugh1998b,McGaugh1998c,McGaugh1999a,McGaugh1999b,McGaugh2000a, McGaugh2000b,McGaugh2001} (and others, variously)   added considerably to the volume of work demonstrating the absolute efficacy of the MOND algorithm in the context of disk galaxies. These references are inclusive up until about the turn of the century.
\\\\
The primary argument levelled against MOND (apart from the fact that its successes appear to be confined to the domain of galaxies) is that it has no theoretical support, although significant effort has been expended in trying to build theories around it. By contrast, neo-MOND consists of a two-component scaling relationship which arises as an exact solution of a special case (one restricted to consider only circular motions) of a general cosmology  rooted in very old ideas that trace via the ideas of Leibniz \& Mach arguably to those of Aristotle. 
The general cosmology concerned, which first appeared in the mainstream literature in a primitive form as \citet{Roscoe2002A} and \citet{Roscoe2004} (together with associated data analyses \citet{Roscoe1999}, \citet{Roscoe2002B}), has evolved sufficiently(arXiv paper \citet{Roscoe2018}) to provide for the present development. The pivotal two-component scaling relationship is made explicit in \S\ref{neo-MOND}, and its development is summarized in appendix \S\ref{Outline}.
\\\\
Neo-MOND is so named primarily because the critical acceleration parameter, $a_0$, of Milgrom's MOND is fundamental to it, and because this parameter is similarly associated with a critical acceleration boundary which, in neo-MOND's case, is the boundary separating the two components of the scaling relationship. However, there is one particularly significant difference separating the two theories: absent from MOND, but fundamental to neo-MOND, is \emph{the MOND acceleration hypothesis}. Because of its simplicity and significance, we describe it in the following.
\subsection{The MOND acceleration hypothesis \& the distance scale} \label{MOND-AC}
In principle, classical MOND contains a discontinuity in the  radial gradient of the rotational velocity $V_{rot}(R)$ at the radius where one gravitational law is considered to give way to another. In practice, this discontinuity  is treated as an artifact of the MOND process and is not considered to have any physical significance for real rotation curves. It is routinely smoothed away using an interpolation function. 
\\\\
Similarly, the two-component scaling relationship which is neo-MOND also contains a discontinuity in the  radial gradient of the rotational velocity $V_{rot}(R)$ at the boundary between the two scaling relations. However, neo-MOND differs from MOND in that it considers this discontinuity to be an unambiguous signature of a fundamental change in the physical regime -  and so it is considered to be a potentially detectable and equally fundamental feature of real rotation curves. So, given that such discontinuities in disks exist and can be objectively detected (which is shown to be the case herein) we are led to:
\begin{quote}
	\emph{The MOND acceleration hypothesis:} For any given disk, the location of the  discontinuity in the radial gradient of the rotation velocity coincides with the location at which the MOND critical gravitational acceleration is reached,
\end{quote}
together with the corollary:
\begin{quote}
	\emph{The linear scale hypothesis:}  The distance scale (and hence the linear scale) for any given disk galaxy is correctly set when the requirements of the MOND acceleration hypothesis are satisfied.
\end{quote}
The support for these linked statements is statistically compelling: assuming circular motions, it amounts to the fact that when  the distance scale (and hence the linear scale) for each object in the SPARC sample is adjusted to guarantee that $V_{rot}^2/R \approx a_0$ at the detected radial gradient discontinuity, then the complete disorder which exists between theoretical mass determinations,  $M$(theory), and SPARC photometric mass determinations, $M$(photometry), gives way to the perfect order of an exact statistical correspondence $M{\rm(theory)}\approx M{\rm(photometry)}$. The algorithmic details for this process are described in \S\ref{SA}, whilst the analysis itself is detailed in \S\ref{Prelim}, \S\ref{SPARCmass1} and \S\ref{SPARCmass2}. A summary of these results is given in the following.
\subsection{Summary of results} \label{Overview}
Briefly, neo-MOND takes the form of a two-component scaling relationship for disk objects, and so has the following structure:
\begin{eqnarray}
S_1(R,V_{rot}(R),\Sigma_R,\Sigma_0,\Sigma_F,V_{flat})&=&0,~~~~~R\leq R_0; \nonumber \\
S_2(R,V_{rot}(R),\Sigma_R,\Sigma_0,\Sigma_F,V_{flat})&=&0,~~~~~R > R_0 \nonumber
\end{eqnarray}
where $\Sigma_R$ is the mass surface density at radius $R$, $\Sigma_0$ is the mass surface density at $R_0$,   $\Sigma_F \equiv a_0/(4\pi G)$ is a characteristic mass surface density parameter associated with the MOND critical acceleration $a_0$ and, by the MOND acceleration hypothesis, the radial gradient discontinuity at $R_0$ is coincident with the MOND critical acceleration radius.
\\\\
The results flowing from this two-component scaling relationship can be listed as:
\begin{itemize}
	\item $S_1(R_0)=0$ is a quantitative refinement of Freeman's Law; 
	\item the BTFR emerges directly when $S_1(R_0)=0$ is constrained by the MOND acceleration hypothesis;
	\item in \S\ref{SPARCmass1} it is shown how the radial distribution of mass on the interior, $M(R \leq R_0)$(theory), tracks SPARC photometry with perfect statistical fidelity, using a global $MLR=2.0$;
	\item in \S\ref{SPARCmass2} it is shown how the radial distribution of mass on the exterior, $M(R > R_0)$(theory) tracks SPARC photometry up to $M_{flat}$ with perfect statistical fidelity, using a global $MLR=2.0$;
	\item in \S\ref{RCbehaviour-1} it is shown how the scaling relationship $S_2(R>R_0)=0$ provides a perfect understanding of why RCs, having reached $(R_0,V_0)$ continue by:
	\begin{itemize}
	 \item either rising smoothly on an asymptotic approach to a rotation velocity $V_{flat} > V_0$;
	 \item or changing abruptly to flatness with a rotation velocity $V_{flat} = V_0$; 
	 \item or falling smoothly on an asymptotic approach to a rotation velocity $V_{flat}< V_0$.
	 \end{itemize}
\end{itemize}
\section{General considerations around neo-MOND} \label{GeneralConsiderations}
The derivation of neo-MOND is sketched out in \S\ref{Outline} but, briefly, it amounts to a local spherically symmetric perturbation of the equilibrium state of a particular cosmology, \citet{Roscoe2002A}, \citet{Roscoe2004},  \citet{Roscoe2018} rooted in the ideas of Leibniz and Mach; according to this equilibrium state, matter in dynamical equilibrium is necessarily distributed in a fractal $D=2$ fashion extending over all scales. 
\\\\
This theoretical result is in accordance with the now accepted reality is that, \emph{on medium scales at least}, matter in the universe is, in a statistical sense, distributed in a  $D\approx2$ quasi-fractal manner. This empirical fact, when taken seriously, proves to be absolutely pivotal to an understanding of MOND phenomonology for a very simple reason: taken seriously, a quasi-fractal world of $D\approx 2$ implies the existence of a characteristic mass surface-density, $\Sigma_F$ say, on the medium scales concerned. An ambient mass surface-density, if you like. The existence of a characteristic acceleration scale, $a_F \equiv 4\pi G \,\Sigma_F$, where $G$ is the gravitational constant, then follows as a matter of course. If one makes the obvious general association that MOND $a_0 \equiv a_F$, then we immediately have
\begin{equation}
\Sigma_F = \frac{a_0}{4\pi G}  \label{eqn1}
\end{equation} 
and a window is opened onto the whole of MOND phenomonology. For this reason, \S\ref{Observations} provides a brief overview of the historical debate surrounding questions of large scale structure.
\\\\
However, there is a further obvious implication which must be considered: as we have implied above, neo-MOND assumes the distribution of material to be $D \approx 2$ quasi-fractal down to scales which include the external local environments of individual disks, and so clearly requires the existence of a non-trivial $D\approx 2$ quasi-fractal intergalactic medium (IGM) having exactly the same properties as the large scale distribution, so that (\ref{eqn1}) applies equally. 
\\\\
We now come to an obvious question: if such an IGM exists, why has it not been detected? The answer to this question depends on the considerations of \S\ref{Observations}  and is addressed in \S\ref{Distances1} \& \S\ref{Distances2}.
\subsection{The external environment: the observations \& the debate}\label{Observations}
A basic assumption of the \textit{Standard Model}
of modern cosmology is that, on some scale, the universe is homogeneous;
however, in early responses to suspicions that the accruing data was
more consistent with Charlier's conceptions \citet{Charlier1908, Charlier1922, Charlier1924}
of an hierarchical universe than with the requirements of the \textit{Standard
	Model},  \citet{De Vaucouleurs1970} showed that, within wide limits,
the available data satisfied a mass distribution law $M\approx r^{1.3}$,
whilst \citet{Peebles1980} found $M\approx r^{1.23}$. The situation,
from the point of view of the \textit{Standard Model}, continued to
deteriorate with the growth of the data-base to the point that, \citet{Baryshev1995}  were able to say
\begin{quote}
	\emph{...the scale of the largest inhomogeneities (discovered to date)
		is comparable with the extent of the surveys, so that the largest
		known structures are limited by the boundaries of the survey in which
		they are detected.}  
\end{quote}
For example, several redshift surveys of the late 20th century, such
as those performed by \citet{Huchra1983}, \citet{Giovanelli1986}, \citet{DeLapparent1988}, \citet{Broadhurst}, \citet{DaCosta} and \citet{Vettolani} 
etc discovered massive structures such as sheets, filaments, superclusters
and voids, and showed that large structures are common features of
the observable universe; the most significant conclusion drawn from
all of these surveys was that the scale of the largest inhomogeneities
observed in the samples was comparable with the spatial extent of
those surveys themselves.\\
\\
In the closing years of the century, several quantitative analyses
of both pencil-beam and wide-angle surveys of galaxy distributions
were performed: three examples are given by \citet{Joyce}  who analysed the CfA2-South catalogue to find fractal
behaviour with $D\,$=$\,1.9\pm0.1$; \citet{SylosLabini}
analysed the APM-Stromlo survey to find fractal behaviour with $D\,$=$\,2.1\pm0.1$,
whilst \citet{SylosLabini1} analysed the
Perseus-Pisces survey to find fractal behaviour with $D\,$=$\,2.0\pm0.1$.
There are many other papers of this nature, and of the same period,
in the literature all supporting the view that, out to $30-40h^{-1}Mpc$
at least, galaxy distributions appeared to be consistent with the simple stochastic fractal model with the critical fractal dimension of $D\approx  D_{crit} = 2$.\\
\\
This latter view became widely accepted (for example, see  \citet{Wu}), and the open question became whether or not
there was transition to homogeneity on some sufficiently large scale.
For example, \citet{Scaramella} analyse the ESO Slice
Project redshift survey, whilst \citet{Martinez} analyse
the Perseus-Pisces, the APM-Stromlo and the 1.2-Jy IRAS redshift surveys,
with both groups claiming to find evidence for a cross-over to homogeneity
at large scales.\\
\\
At around about this time, the argument reduced to a question of
statistics (\citet{Labini}, \citet{Gabrielli}, \citet{Pietronero}):
basically, the proponents of the fractal view began to argue that
the statistical tools (that is, two-point correlation function methods) widely used
to analyse galaxy distributions by the proponents of the opposite
view are deeply rooted in classical ideas of statistics and implicitly
assume that the distributions from which samples are drawn are homogeneous
in the first place.  \citet{Hogg}, having accepted
these arguments, applied the techniques argued for by the pro-fractal
community (which use the \emph{conditional density} as an appropriate
statistic) to a sample drawn from Release Four of the Sloan Digital
Sky Survey. They claimed that the application of these methods does
show a turnover to homogeneity at the largest scales thereby closing,
as they see it, the argument. In response, \citet{SylosLabini2} 
criticized their paper on the basis that the strength of the
conclusions drawn is unwarrented given the deficencies of the sample
- in effect, that it is not big enough. 
\\\\
More recently, \citet{Tekhanovich} have addressed the deficencies of the Hogg et al analysis by analysing the 2MRS catalogue, which provides redshifts of over 43,000 objects out to about 300Mpc, using conditional density methods; their analysis shows that the distribution of objects in the 2MRS catalogue is consistent with the simple stochastic fractal model with the critical fractal dimension of $D\approx  D_{crit} = 2$.
\\\\
To summarize, the proponents of non-trivially fractal large-scale
structure have won the argument out to medium distances and the controversy
now revolves around the largest scales encompassed by the SDSS.
\subsection{General properties of a $D\approx 2$ fractal Intergalactic Medium}\label{Distances1}
Because of a perceived and unexplained over-dimming of type-SN1a supernovae, it has been hypothesized for some considerable time that there exists a non-trivial intergalactic medium (IGM) consisting of \lq{grey dust}' (dust which causes extinction, but very little reddening) expelled from the galaxies, to account for this. For example, see \citet{AA1999} and  \citet{PSC2006}. The principle of a \lq{grey dust}' IGM in some form or other is thus established, albeit as a means of explaining anomolous dimming of SN1a objects at high redshifts. Because of the high redshifts involved here, the inherent assumption is that the grey dust involved exists at extremely low column densities.
\\\\
However,  we can note that whilst the various proposed models assume some degree of homogeneity in the distribution of grey dust, the considerations of \S\ref{Observations} suggest otherwise: specifically, since galaxies on the medium distance scale are observed to be distributed in a $D\approx 2$ quasi-fractal manner, then so must any grey dust which originates in them be likewise distributed.
By virtue of its $D\approx 2$ quasi-fractal distribution, such an IGM would, to a significant extent, be transparent to radiation, which mirrors the primary reason why \citet{Charlier1908} suggested  the \lq{hierarchical universe}' as a possible answer to the question \emph{Why is the sky dark at night?}
\\\\
To complicate the issue even further, we can reasonably suppose that this grey dust (should it exist) would be in thermodynamic equilibrium with the cosmic background, and therefore difficult, if not impossible, to distinguish from it.
The net result of these considerations is that an IGM consisting of a $D\approx 2$ quasi-fractal distribution of grey dust would  be extremely difficult to detect directly.
\subsection{Observational consequences of a  $D\approx 2$ fractal grey dust IGM} \label{Distances2}
Whilst a $D\approx 2$ quasi-fractal grey dust IGM would, to a significant extent, be transparent to radiation,
it would by no means be totally transparent and, broadly speaking, light from a source at distance $R$ would dim (without reddening) via a process of extinction by the grey dust in a way which would be proportional to $R^2$, again because this material is distributed quasi-fractally, $D\approx 2$. 
\\\\
In the absence of reddening, this dimming mechanism would be indistinguishable in its effects from the ordinary inverse-square distance dimming process so that the total of observed dimming would be interpreted entirely as a distance effect. There are two consequences:
\begin{itemize}
	\item The principles underlying the process by which standard candles are used to estimate the absolute luminosities of distant objects are unchanged so that such estimates would not be affected by grey dust extinction, should the phenomenon actually exist;
	\item The photometric distance scale would be systematically exaggerated with the effect that objects of a given absolute luminosity would generally be estimated as being further away and larger than they actually are. In passing, we can see that, in the context of classical gravitational theory, such an exaggeration of the distance scale would automatically give rise to a \lq{missing mass}' problem.
\end{itemize}
It is to be emphasized that it would be extremely difficult, if not impossible, to detect such a grey dust IGM by any direct means. Any dimming arising from its presence would be interpreted entirely as a distance effect.
\section{MOND and neo-MOND}
In this section, we give a very brief overview of MOND and of neo-MOND. The irreducible connection between the two is simply that the critical gravitational acceleration scale
\begin{equation}
a_0 \approx 1.2\times 10^{-10} m/sec^2, \label{eqn1A}
\end{equation}
is fundamental to both.  However, apart from the fact that neo-MOND arises from an underlying theory in which all conservation laws are satisfied, its interpretation is quite different from that of MOND. The primary difference is that, whereas for MOND, the parameter $a_0$ defines the critical gravitational acceleration scale at which one gravitational law gives way to another, for neo-MOND  this critical acceleration scale signifies the boundary $R=R_0$ between two distinct physical regimes, each with its own distinct scaling laws: one the galactic interior contained within the critical acceleration radius; the other the $D\approx 2$ quasi-fractal IGM with a characteristic mass surface density $\Sigma_F = a_0/(4\pi G)$.
\subsection{Milgrom's classical MOND}
The MOND algorithm (which, in its original form, assumed purely circular motions) is built around the idea that for gravitational accelerations $ > a_0$ then classical Newtonian physics applies, whilst for gravitational accelerations $ \leq a_0$, then the provisions of MOND apply.
To be specific, in it's original formulation as a modification of gravity (rather than of inertia),  the algorithm (\citet{Sanders2002A}) relates the actual gravitational acceleration, $\mathbf{g}$ say, to the Newtonian gravitational acceleration, $\mathbf{g}_N$ say,
via the relation
\[
\mathbf{g}\, F(g /a_0) = \mathbf{g}_N,
\]
where $g \equiv \left| \mathbf{g}\right|$ and $F(x)$ is some heuristically determined function satisfying $F(x) \rightarrow 1$ for $x  \rightarrow {\rm large}$ and $F(x)\rightarrow x$ for $x \rightarrow {\rm small}$. For $g << a_0 $, this implies
\[
g = \sqrt{a_0 \,g_N}
\]
so that in the case of a point source of mass $M$ and purely circular velocities of magnitude $V$, we have the flat rotation velocity given by
\begin{equation}
V_{flat} = \left( a_0\, G\, M \right)^{1/4} \label{eqn1a}
\end{equation}
which, of course, is a very specific quantitative form of the baryonic Tully-Fisher relation (BTFR). 
\\\\
In the more general case of an object with an extended mass distribution, for which $M\equiv M_{flat}$ is re-interpreted as the amount of \emph{visible} mass detectable up to the radius at which $V_{flat}$ is considered attained, (so that no notion of DM is entailed), then (\ref{eqn1a}), written as
\begin{equation}
V_{flat} = \left( a_0\, G\, M_{flat} \right)^{1/4}\,, \label{eqn1b}
\end{equation}
is absolute for Milgrom's MOND in all circumstances, irrespective of the details of the mass distribution in the object concerned.
\\\\
For the determination of $M_{flat}$, it is generally assumed that good estimates of the mass distributions of gas and dust in disks can be determined from their radiative properties alone, and that the only significant uncertainties concern stellar mass-to-light ratios in particular. Consequently, the only free parameter within the MOND algorithm is the stellar mass-to-light ratio, $\Upsilon$. For any given disk, $\Upsilon$ is varied to ensure that the correct asymptotic rotation velocity is obtained for the disk concerned. Remarkably, all the other details of the rotation curve then fall automatically into place.
\subsection{Neo-MOND}\label{neo-MOND}
Neo-MOND arises as a modelling exercise (appendix \S\ref{Outline}) from a general cosmology  within which all conservation laws are satisfied - see arXiv paper \citet{Roscoe2018}, which is the latest evolution of work originally published, in a primitive form, in the mainstream literature as  \citet{Roscoe2002A}. A basic modelling assumption of neo-MOND (but not of the general theory), shared with classical MOND in its early applications to disk galaxies, is that all motions are circular; under this assumption, the general theory admits a denegerate state with an associated energy equation, (\ref{eqn4g}), which quickly leads to the following scaling relations for the internal dynamics of a simple disk galaxy sitting within an external environment which is assumed to be primarily gaseous nearby merging into the a \lq{grey dust}'  $D\approx 2$ quasi-fractal IGM:
\begin{eqnarray}
\frac{V_{rot}(R)}{V_{flat}} &=& \left(\frac{ \Sigma_F }{   \Sigma_R } \right)^{1/2}, ~~~~ R \leq R_0; \nonumber \\
\label{eqn2} \\
\frac{V_{rot}(R)}{V_{flat}} &=&  \left(\frac{ \Sigma_F  R^2}{  \left(\Sigma_0 - \Sigma_F\right)R_0^2 + \Sigma_F R^2 } \right)^{1/2},~~~~ R > R_0. \nonumber 
\end{eqnarray}
The primary disk parameters in these scaling relations are $(R_0, \Sigma_0, V_{flat})$ where:
\begin{itemize}
	\item $R_0$, the critical radius, is the radius at which the radial gradient of $V_{rot}(R)$ is discontinuous, marking the boundary between two, quite distinct, physical regimes; 
	\item $R_0$ is hypothesized to coincide with the radial position at which the MOND critical acceleration $a_0$ is reached - this is \emph{the MOND acceleration hypothesis};
	\item $\Sigma_R \equiv \mathcal{M}_g(R)/(4\pi R^2)$ is the mass surface density at $R \leq R_0$, where $\mathcal{M}_g(R\leq R_0)$ represents the radial distribution of mass within the critical radius;
	\item $\Sigma_0 \equiv \mathcal{M}_g(R_0)/(4 \pi R_0^2)$ is the mass surface density at $R_0$;
	\item $\Sigma_F \equiv a_0/(4\pi G)$ is interpreted as the characteristic mass surface density of the  $D\approx 2$ quasi-fractal IGM;
	\item $V_{flat}$ is the flat rotation velocity.
\end{itemize} 
\subsection{LSB objects in neo-MOND}
At face value, it would appear that the rotation curves of LSBs could not possibly be modelled by (\ref{eqn2}) simply because such objects are defined as ones for which the whole disk is in the low acceleration MOND regime, implying $R_0=0$ for such disks, and hence $V_{rot}(R>0) = V_{flat}$. We address this objection explicitly in \S\ref{LSBs} for the 25 SPARC objects identified as LSBs in the source papers used by \citet{McGaugh2015}. 
\section{A refined Freeman's Law \& the BTFR} \label{BTFR}
In the following, we show how a quantitative refinement of Freeman's Law arises directly as a special case of the neo-MOND scaling relationships and how, when this is constrained by the MOND acceleration hypothesis, the BTFR emerges naturally.
\subsection{The refined Freeman's Law } \label{FL-1}
From (\ref{eqn2}), at the radial gradient discontinuity $R=R_0$ we have immediately 
\begin{equation}
\left(\frac{V_0}{V_{flat}}\right)^2 = \frac{\Sigma_F}{\Sigma_0}  \label{eqn4e}
\end{equation}
from which we get an obvious refinement of Freeman's Law 
\begin{equation}
\Sigma_0 =\left( \frac{V_{flat}}{V_0}\right)^2  \Sigma_F  \label{eqn4f}
\end{equation}
relating the mass surface density of the galaxy at the critical radius to the characteristic mass surface density of the $D\approx 2$ fractal grey dust IGM.
\\\\
This general form makes it clear that the classical statement of Freeman's Law is restricted to those objects for which $V_0=V_{flat}$; that is, to those objects for which the critical acceleration, $a_0$, coincides with an abrupt transition to flatness. Such objects are those with architypal flat rotation curves - within the SPARC sample, two clear examples of such objects are ESO563-G02 and  NGC2998. 
\subsection{The BTFR} \label{BTFR-1}
From (\ref{eqn4e}) and using (\ref{eqn1}), we have 
\begin{equation}
V_0^2 = V_{flat}^2 \,\frac{\Sigma_F}{\Sigma_0} =  V_{flat}^2 \left( \frac{a_0 R_0^2}{G M_0}\right) \label{eqn4ee}
\end{equation} 
where $M_0 \equiv \mathcal{M}_g(R_0)$ is the mass contained within $R=R_0$.
If we now constrain this by the MOND acceleration hypothesis, then directly: 
\begin{equation}
\frac{V_0^2}{R_0} =  V_{flat}^2 \left( \frac{a_0 R_0}{G M_0}\right) = a_0, \label{eqn3NN}
\end{equation}
which is the quantitative form of the acceleration hypothesis. Eliminating $R_0$ between these two equations gives directly:
\begin{equation}
V_{flat}^4 = a_0\, G\, \left[ \left(\frac{V_{flat}}{V_0} \right)^2 M_0 \right].  \label{eqn4d}
\end{equation}
Defining $M_{flat}$(theory) according to
\begin{equation}
M_{flat}{\rm(theory)} \equiv \left(\frac{V_{flat}}{V_0} \right)^2 M_0, \label{eqn5d}
\end{equation}
then (\ref{eqn4d}) becomes
\begin{equation}
V_{flat}^4 = a_0\, G\, M_{flat}{\rm(theory)}  \label{eqn5c}
\end{equation}
which has the exact structure of Milgrom's form (\ref{eqn1b}) of the empirical BTFR. So everything hinges on the extent to which $M_{flat}{\rm(theory)}$ tracks $M_{flat}$(photometry), where we remember that $M_{flat}$(photometry) is conventionally defined as the photometrically estimated mass contained up to $V_{flat}$. In practice, as is shown in \S\ref{Mflat},  we find $M_{flat}{\rm(theory)} = M_{flat}{\rm(photometry)}$ at the level of statistical certainty over the SPARC sample.
\\\\
However, notwithstanding the quality of this statistical result, it is clear that in those few cases for which $V_0>V_{flat}$, then  (\ref{eqn5d}) explicitly states that $M_{flat}{\rm(theory)} < M_0$. Consequently, since it is certainly true that $M_0<M_{flat}$(photometry), then  $M_{flat}{\rm(theory)} = M_{flat}{\rm(photometry)}$ cannot be true on the conventional definition of $M_{flat}{\rm(photometry)}$ in these few cases. The questions raised by this result are resolved in detail in the following,  \S\ref{SpecialCase-1}.
\subsection{The special case: $V_0 > V_{flat}$: A finite disk-boundary} \label{SpecialCase-1}
This is a minority case for the objects of the SPARC sample and crucially by (\ref{eqn4f}) (the refined Freeman's Law) is associated with the mass surface density condition $\Sigma_0 < \Sigma_F$. The considerations of \S\ref{RCbehaviour-1} show  that the RCs of such objects are characterized by a smooth \emph{descent} from $V_{max} \equiv V_0 > V_{flat}$ to approach $V_{flat}$ asymptotically.
\\\\
This case creates an interesting question in that, according to (\ref{eqn5d}), it implies $M_{flat}{\rm(theory)} < M_0$
so that $M_{flat}$ cannot be the contained luminous mass at the asymptotic $V_{flat}$ rotation velocity, contrary to the photometric definition. However, in such cases, the velocity value $V_{flat}$ is reached \emph{twice} on the rotation curve - once on the rising part before $V_{max}\equiv V_0$ is reached, and once asymptotically after $V_{max} \equiv V_0$ is reached.
\\\\
It follows that (\ref{eqn5d}) can be consistently interpreted in this special case if $M_{flat}$ is understood to be the contained luminous mass up to the \emph{first} occurrence of $V_{flat}$ on the rotation curve. It is this value of $M_{flat}$ which satisfies the BTFR in the theoretical development of \S\ref{BTFR}. It is clear that $M_{flat}{\rm(theory)} < M_{flat}{\rm(photometry)}$ for this case but, in practice, we find that the corrections required to make  the definition of $M_{flat}$(photometry) consistent  are too small to make any noticeable difference to the analysis of  \S\ref{Mflat}.
\section{The neo-MOND modelling process} \label{SA} 
Classical MOND uses   photometrically determined mass-modelling within galaxy disks, in conjunction with photometrically determined distance scales, to predict the details of the associated rotation curves. Neo-MOND is applied in the reverse way: we use dynamical modelling of the rotation curves to determine the absolute distance scales, characteristic mass parameters $(M_0, M_{flat})$, and the total radial distribution of disk mass up to $M_{flat}$  for each disk, which we then compare against the photometric estimations (SPARC photometry) of the same quantities. 
\subsection{The algorithmic details}\label{Details}
The mass surface density function, $\Sigma_R$ of (\ref{eqn2}), is given by 
\[
\Sigma_R \equiv \frac{\mathcal{M}_g(R)}{4 \pi R^2},~~~ R \leq R_0
\]
where $\mathcal{M}_g(R)$ is the mass contained within radius $R$. This latter function is not yet specified.
However, in the general theory from which neo-MOND is derived, the use of the simple model 
\begin{equation}
\mathcal{M}_g(R) \equiv M_0 \left(\frac{R}{R_0} \right), ~~R \leq R_0, \label{eqn8}
\end{equation}
reduces that general theory exactly to Newtonian gravitation on $R \leq R_0$.  Since Newtonian gravitation is known to work tolerably well on the inner regions of galaxy disks, this linear mass model is used exclusively from now on. Note that $M_0 \equiv \mathcal{M}_g(R_0)$.
\\\\
Taking note of some essential computational details described in \S\ref{MassModels}, the neo-MOND scaling relations of (\ref{eqn2})  can now be applied as follows:
\begin{enumerate}
\item  The modelling algorithm treats the three parameters $( R_0, \Sigma_0,  V_{flat})$ as independent, all to  be varied in order to optimize the fit of $V_{rot}(R)$ given at (\ref{eqn2}) to the SPARC rotation curve in the disk concerned. An automatic code, based on the Nelder-Mead method (robust on noisy data), is used for this process;
\item Compute $M_0=4 \pi R_0^2\, \Sigma_0$, then compute $V_0$ from (\ref{eqn4e}), and finally compute $M_{flat}$ from (\ref{eqn5d}). The full set of characteristic parameters $(R_0, V_0, M_0, V_{flat}, M_{flat})$ computed according to the photometric distance scalings implicit to SPARC for the given disk are then available;
\item At this stage, all computations have been done assuming the distance scales implicit to the SPARC sample so that, generally speaking, the requirements of the MOND acceleration hypothesis are not met. In fact, we routinely find
\[
M_0({\rm theory}) >> M_0({\rm photometry}),~~M_{flat}({\rm theory}) >> M_{flat}({\rm photometry})
\] 
so that, as with conventional theory, a \lq{missing mass}' problem has emerged; 
\item But neo-MOND requires the conditions of the MOND acceleration hypothesis to be satisfied. So, rescale each disk according to $R\rightarrow  K R$, where $K$ is chosen to ensure that (\ref{eqn3NN}) is exactly satisfied; 
\item From (\ref{eqn2}), when $R$ is rescaled in this way, then the calculated values of $M_0$ must also be rescaled according $M_0\rightarrow  K^2 M_0$ in order to ensure that $\Sigma_0$ and hence the velocities remain invariant. Since $M_0$ has been rescaled, then (\ref{eqn5d}) indicates that $M_{flat} \rightarrow K^2 M_{flat}$ also;
\item A comparison with SPARC photometry in \S\ref{M0} and \S\ref{Mflat} respectively now shows:
\[
M_0({\rm theory}) = M_0({\rm photometry})~~\&~~M_{flat}({\rm theory}) = M_{flat}({\rm photometry})
\] 
 at the level of statistical certainty over the whole of the SPARC sample. There is no mass discrepancy;
\item We can now compute the radial distribution of disk mass over $R \leq R_0$ using $\mathcal{M}_g(R\leq R_0)$ defined at (\ref{eqn8}) to find: 
 \[
 M(R \leq R_0)({\rm theory}) = M(R \leq R_0)({\rm photometry}),
 \] 
at the level of statistical certainty over the whole of the SPARC sample. The details are shown in \S\ref{Interior}. There is no mass discrepancy;
 \item Using an obvious generalization of the scaling relation (\ref{eqn5d}), we can similarly compute the radial distribution of disk mass over $R>R_0$ up to $M_{flat}$ to find:
 \[
 M(R)({\rm theory}) = M(R)({\rm photometry}),~~~R > R_0,~~~ M(R) \leq M_{flat}
 \] 
 at the level of statistical certainty over the whole of the SPARC sample. The details are shown in \S\ref{Exterior}. There is no mass discrepancy.
\end{enumerate} 
 As we shall see in \S\ref{SPARCmass1} and \S\ref{SPARCmass2}, the statistical power of these results represents 
 compelling evidence in support of the MOND acceleration hypothesis and its corollary, the distance scale hypothesis of \S\ref{MOND-AC};
\subsection{Dealing with noisy data}
In general, the foregoing requires that we ask whether the relationship 
\begin{equation}
Mass({\rm theory}) \approx Mass(\rm{photometric}) \label{Mass1}
\end{equation}
is supported on the data.
As we shall see, the correlation between these two quantities, after rescaling using the MOND acceleration hypothesis (but not before), is statistically extremely powerful.
\\\\
However, notwithstanding this powerful correlation,  both quantities are very noisily determined, and so the standard tool of least-squares linear regression is particularly poorly suited for the task of determining any quantitative relationship between them, since this standard tool assumes the predictor to be entirely free of error. In consequence, any quantitative relationship we deduce from  using least-squares linear regression is always strongly dependent on the choice of the predictor (which one is assumed to be \lq{most}' error-free), and therefore cannot be relied upon in the present context.
\\\\
We eliminate this problem by developing a method of linear regression based on \emph{least areas}, which treats predictor data and response data in an entirely symmetric fashion. Consequently, the quantity being minimised is independent of how the predictor/response pair is chosen for the regression. The result is a linear model which can be algebraically inverted to give the exact linear model which would also arise from regressing on the interchanged predictor/response pair. So, any inference drawn about the relationship between the predictor and response is independent of how the predictor/response pair is chosen. 
The details of the derivation are given in \S\ref{LeastAreas}, and a brief description of least-area linear regression is given below.
\\\\
For a model $y = A x + B$ fitted to the data $(X_i, Y_i), \,i=1 ... N$,  least-areas regression defines the parameters according to: 
\[
A = \sqrt{\frac{\left( \Sigma Y_i\right)^2 - N \Sigma Y_i^2}{\left( \Sigma X_i\right)^2 - N \Sigma X_i^2}}, ~~~~ B = \frac{\Sigma Y_i - A \Sigma X_i}{N}.
\]
To see the algebraic invertability property,  fit the model $x = \alpha y + \beta$ to the same data, and we get:
\[
\alpha = \sqrt{\frac{\left( \Sigma X_i\right)^2 - N \Sigma X_i^2}{\left( \Sigma Y_i\right)^2 - N \Sigma Y_i^2}}, ~~~~ \beta = \frac{\Sigma X_i - \alpha \Sigma Y_i}{N}.
\]
Comparing the two models quickly shows that $\alpha = 1/A$ and $\beta = -B/A$ so that $x=\alpha y + \beta$ is the algebraic inverse of $y = A X + B$.
\section{Theory against observation: preliminary comments} \label{Prelim}
We consider only the SPARC galaxies which have quality flag Q = 1 or 2, which gives a total sample of 160 objects out of a total of 175 objects. For simplicity, and ease of making the main points, we then select only those objects which appear to have no measurable bulge component (that is, are explicitly stated to have no measurable bulge component in the database), giving a final sample of 129 objects.
\\\\
Given that a fixed MLR of 2.0 is applied to the photometry across the whole SPARC sample, then the only parameters varied in the algorithm of \S\ref{Details} to optimize RC fits are the three characteristic disk parameters $(R_0, \Sigma_0, V_{flat})$ from which, using (\ref{eqn4e}), (\ref{eqn5d}) and $M_0=4\pi R_0^2\,\Sigma_0$, we obtain the full set of characteristic disk parameters $(R_0, V_0, M_0, V_{flat}, M_{flat}{\rm(theory)})$. 
\\\\
Having calculated $M_0$ in \S\ref{M0} it becomes possible to calculate the radial distribution of luminous mass on $R < R_0$ which is described in \S\ref{Interior}. Similarly, having calculated $M_{flat}$(theory) in  \S\ref{Mflat} it becomes possible to calculate the radial distribution of luminous mass on $R > R_0$ up to  $M_{flat}$(theory) which is described in \S\ref{Exterior}. 
\\\\
These four calculations provide quantitative estimations of the luminous mass distribution across the whole disk of each object in the sample, derived purely from disk dynamics.
\subsection{SPARC LSBs}\label{LSBs}
Before considering the results against SPARC photometry  in detail, it is germane to recognize that, according to the notes in the source papers used by \citet{McGaugh2015} to compile the SPARC sample, that sample contains at least 25 LSB objects, listed as:  F563-1, F568-1, F568-3, F568-V1,  F571-V1, F571-8, F574-1, F583-1, F583-4, NGC3917, NGC4010,  UGC00128, UGC01230, UGC05005,  UGC5750,  UGC05999, UGC06399, UGC06446, UGC06667,  UGC06818,  UGC06917,  UGC06923, UGC06930, UGC06983, UGC07089.
\\\\ 
Since such objects are defined as ones in which the entire disk is in the MOND weak-gravity regime then, for these objects, $R_0=0$,  meaning that, at face value, the MOND acceleration hypothesis cannot be used to scale these objects. For this reason, in the figures \ref{fig:SPARCMASSvsTheoryMass} \& \ref{fig:SSM-4} of \S\ref{SPARCmass1} and figures \ref{fig:SSM-2} \& \ref{fig:SSM-5} of \S\ref{SPARCmass2} we identify these LSB objects as filled red circles so that their behaviour under the analysis can be observed. 
\\\\
Against expectation, and as is absolutely clear from the figures, we find that the LSB objects behave exactly as the non-LSB objects under the MOND acceleration hypothesis.   In practice, this means that, in fact, the objects classified as LSBs in the SPARC sample all have an objectively detected non-trivial physical discontinuity in the radial gradient of their rotation curves which has allowed the automatic \& successful application of the MOND acceleration hypothesis to the objects concerned. In other words, none of the listed objects is an LSB in the sense of the standard definition; the conflict is consistent with the idea that there is an unrecognized mechanism causing the standard photometric distances for these objects to exaggerate the actual distances, giving rise to the appearance that the disks concerned are all in the deep MOND regime. 
\section{Theory vs SPARC photometry: $R\leq R_0$}\label{SPARCmass1}
In the following, the photometric data is from \citet{McGaugh2015}, and the sample consists of all 129 non-bulgy objects with quality flag $Q=1,2$ listed by these authors. The data for NGC5371 is missing since the RC fit for this object failed to converge.
\\\\
All error bars given in the least-area linear regressions are set at the conventional two standard deviations, determined by a bootstrapping process.
\subsection{$M_0 \equiv \mathcal{M}_g(R_0)$\,: the total mass inside the critical radius, $R = R_0$}\label{M0}
The parameter $M_0$ represents the predicted total mass inside the critical radius $R = R_0$ for the galaxy concerned. To obtain the SPARC photometric estimate of the same quantity we simply adopted a global mass-to-light ratio in the disks of $MLR = 2.0$, and then integrated the disk photometry over all $R \leq R_0$ (taking care to use original SPARC scalings) to obtain the photometric estimates of $M_0$. Although this represents a very crude way of representing the contributions of the various components of mass within $R = R_0$, it is very effective for current purposes. The results  for the whole SPARC $Q=1,2$ sample are displayed in the two panels of Figure\,\ref{fig:SPARCMASSvsTheoryMass} and discussed below. 
\subsubsection*{Fig \ref{fig:SPARCMASSvsTheoryMass} Upper: Distance scales from conventional photometry}
\begin{figure}[H]
	\centering
	\includegraphics[width=0.7\linewidth]{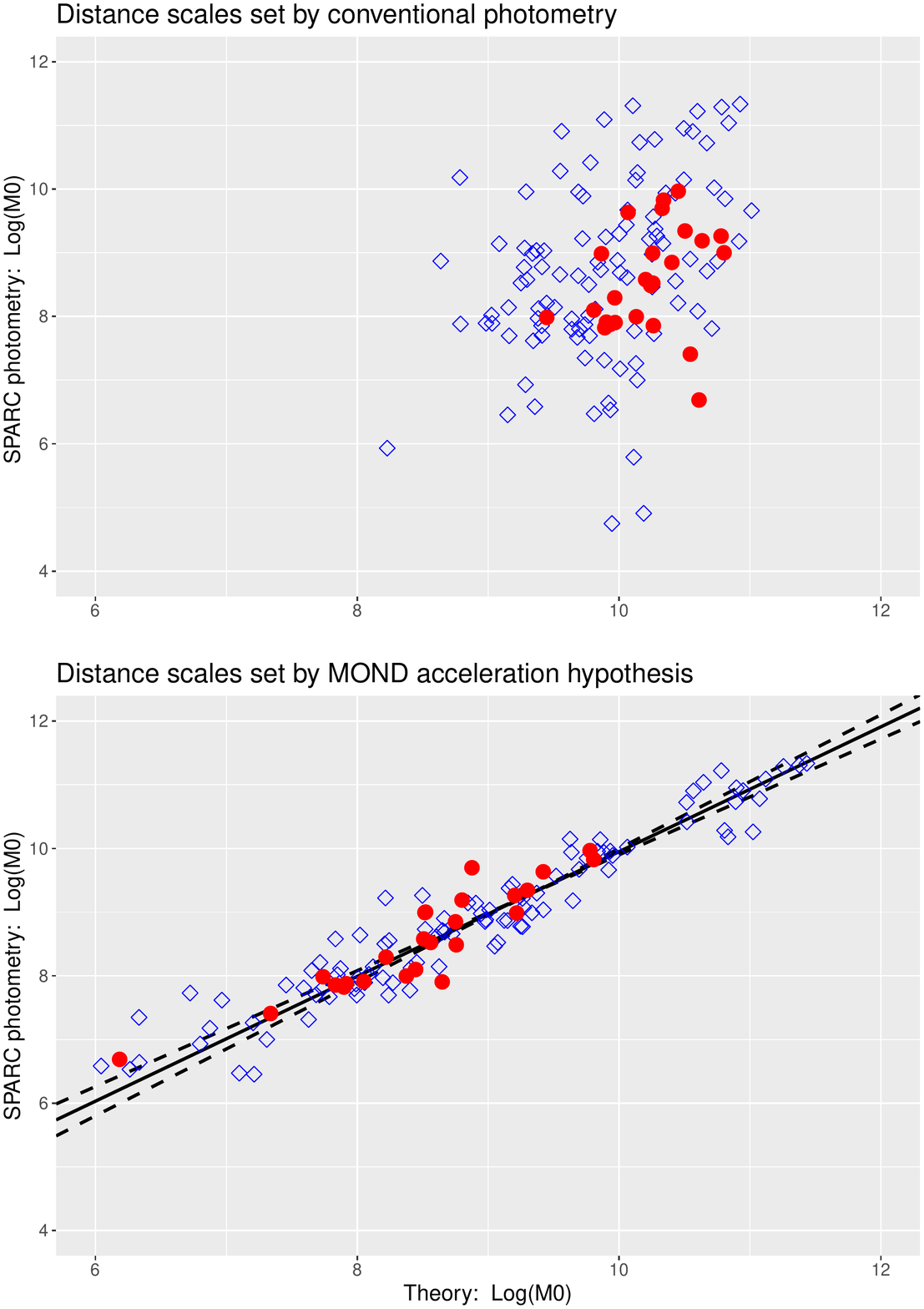}
	\caption{ Red filled circles = putative LSB objects. Open diamonds = everything else. Upper figure:  The $\log{M_0}$(theory) values have been computed on the conventional photometric scalings implicit to the SPARC sample. Lower figure: The upper panel $\log{M_0}$(theory) values have been rescaled according to the requirements of the MOND acceleration hypothesis.}
	\label{fig:SPARCMASSvsTheoryMass}
\end{figure}

Here, putative LSB objects are identified as red filled circles, and $M_0$(theory) is computed by the algorithm of \S\ref{Details} using the photometric distance scalings implicit to the SPARC sample.
The  scatter plot for this case makes it clear that for the most massive objects, say $10 \leq M \leq 12$, there is qualitative agreement between theory and photometry. However, at the low photometric end, we see that $M{\rm(theory)}> M{\rm(photometry)}$ by up to two orders of magnitude. That is, a \lq{missing mass}' problem of similar proportions to that recognized in classical gravitation theory has emerged.
\subsubsection*{Fig \ref{fig:SPARCMASSvsTheoryMass} Lower:  Distance scales from MOND acceleration hypothesis}
Here, putative LSB objects are again identified as red filled circles and $M_0$(theory) computed above has been rescaled according to the requirements of the MOND acceleration hypothesis (\ref{eqn3NN}), specified in \S\ref{Details}.
The  scatter plot for this case makes it clear that this rescaling process has the effect of mapping every data point almost perfectly onto the $\log M_0$(theory) $=\log M_0$(photometry) line. That is, there is now a statistically perfect correspondence between the two mass quantities and a least-area linear regression gives:
\[
\log{M_0}({\rm photometry}) \approx \left(0.99 \pm 0.07\right) \log{M_0}({\rm theory}) + (0.14 \pm 0.62).
\]
In other words, the imposition of the conditions required by the MOND acceleration hypothesis has the effect of imposing an almost perfect correspondence  $M_0{\rm(theory)} \approx M_0{\rm(photometry)}$, so that the \lq{missing mass}' problem of the upper panel has disappeared completely. Note that the behaviour of the  LSB objects cannot be distinguished from that of the non-LSB objects. We have discussed this detail in \S\ref{LSBs}.
\subsection{The radial distribution of mass on the interior: $R < R_0$} \label{Interior}
Using a fixed MLR=2.0, we have integrated the photometry for each disk in the SPARC sample to obtain estimates of $M(R_i)(\rm{photometry})$ contained within radius $R_i$, for each radial coordinate $R_i < R_0$ on the measured disk; this gives a total of 406 individual estimated mass values over the whole non-bulgy sample. In the following, we compare these against theoretical estimates of the same quantities given by the neo-MOND algorithmic process defined in \S\ref{Details}, using the mass model of (\ref{eqn8}).
\subsubsection*{Fig \ref{fig:SSM-4}  Upper: Distance scales from conventional photometry}
\begin{figure}[H]
	\centering
	\includegraphics[width=0.7\linewidth]{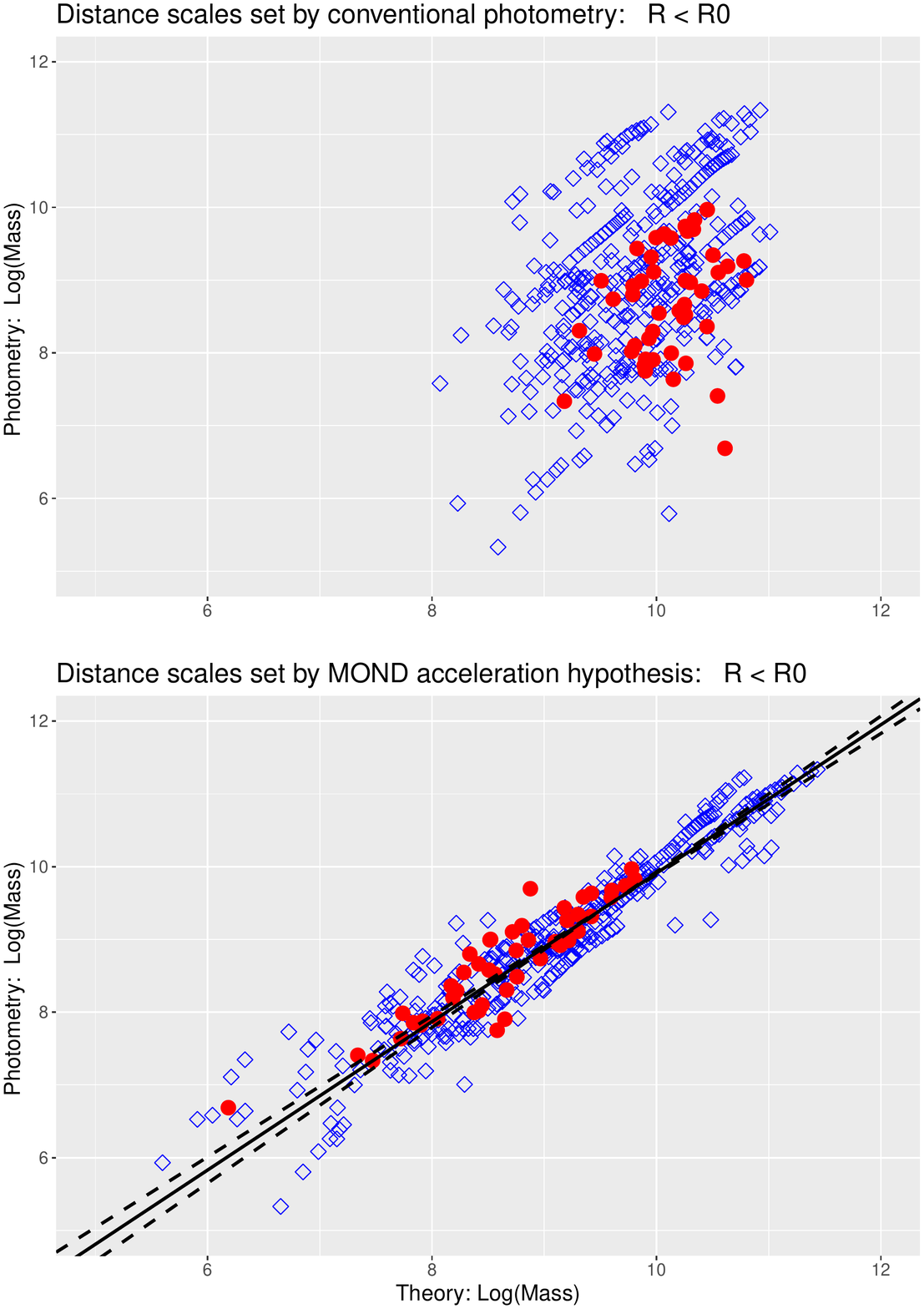}
	\caption{ Red filled circles = putative LSB objects. Open diamonds = everything else. Upper figure:  The  $\log{M(R)}$(theory) values for all disk data points satisfying  $R \leq R_0$ (406 individual points in total) have been computed using the conventional photometric distance scalings implicit to the SPARC sample.  Lower figure: The 406 $\log{M(R)}$(theory) values of the upper panel  have been rescaled according to the requirements of the MOND acceleration hypothesis. }
	\label{fig:SSM-4}
\end{figure}

Here, putative LSB objects are identified as red filled circles and $M$(theory) is computed by the algorithm of \S\ref{Details} on the photometric distance scalings implicit to the SPARC sample. The  scatter plot for this case makes it clear that for the most massive objects, say $10 \leq M \leq 12$, there is qualitative agreement between theory and photometry. However, at the low photometric end, we see that $M{\rm(theory)}> M{\rm(photometry)}$ by up to two orders of magnitude. That is, a \lq{missing mass}' problem of similar proportions to that recognized in classical theory has emerged once more.
\subsubsection*{Fig \ref{fig:SSM-4} Lower: Distance scales from MOND acceleration hypothesis}
Here, putative LSB objects are identified as red filled circles and $M$(theory) computed above have been rescaled according to the requirements of the MOND acceleration hypothesis, specified in \S\ref{Details}.
The  scatter plot for this case makes it clear that, in broad terms, the rescaling process has the bulk effect of shifting $\log{M}({\rm photometry}) < 10.0$ objects to the left in the plot to give an obvious \& powerful correlation. A least-area linear regression gives:
\begin{eqnarray}
\log{M}({\rm photometry}) &\approx& \left(1.03 \pm 0.05\right) \log{M}({\rm theory}) - (0.31 \pm 0.49),
\nonumber 
\end{eqnarray}
som that $\log M(\rm{theory}) \approx \log M(\rm{photometry})$ is confirmed for all the mass measurements inside the critical radius, $R \leq R_0$.
In other words, the imposition of the conditions required by the MOND acceleration hypothesis has the effect of making the \lq{missing mass}' of the upper panel disappear completely. 
\section{Theory vs SPARC photometry: $R > R_0$}\label{SPARCmass2}
In the following, the photometric data is from \citet{McGaugh2015}, and the basic sample consists of all 129 non-bulgy objects with quality flag $Q=1,2$ listed by these authors. The data for NGC5371 is missing since the RC fit for this object failed to converge.
\\\\
All error bars given in the least-area linear regressions are conventionally set at two standard deviations, determined by a bootstrapping process.
\subsection{The distinction between $R\leq R_0$ and $R>R_0$}
There is a fundamental distinction between what happens on $R\leq R_0$ and what happens on $R>R_0$, which must be carefully accounted for.
\\\\
The process of deriving the neo-MOND scaling relationships in \S\ref{Outline} required the core galactic object, represented by the mass distribution $\mathcal{M}_g(R\leq R_0)$, to be explicitly placed into  an unbounded $D=2$ fractal material background as follows: 
\begin{eqnarray}
\mathcal{M}(R) &=& \mathcal{M}_g (R),~~~ R \leq R_0; \nonumber \\
\mathcal{M}(R) &=&  \mathcal{M}_g(R_0) + 4 \pi \Sigma_F \left(R^2 - R^2_0 \right),~~~ R > R_0 \nonumber
\end{eqnarray}
where $\mathcal{M}(R)$ represents the radial distribution of total mass on $0<R<\infty$. 
This theoretical structure is intended to model a galactic object sitting in a hypothetical $D \approx 2 $ quasi-fractal grey dust IGM which, for the reasons discussed in \S\ref{Distances1}, would be extremely difficult, if not impossible, to detect.
\\\\
The distinction between events on $R\leq R_0$ and events on $R>R_0$ is now obvious: In \S\ref{SPARCmass1} it was shown that all of the core object mass represented by $\mathcal{M}_g(R\leq R_0)$ was photometrically accounted for. But we know that on $R>R_0$, the total of all photometrically detectable mass, $M_{flat}$(photometry) say, is finite and bounded, whereas the total mass available on $R>R_0$ is unbounded.
\\\\
So, given, as we shall show, that the scaling relationships on $R>R_0$ actually do account (in a statistically perfect manner) for  the photometrically detectable mass only (and not the generally available mass) we must ask \emph{what is going on?}   
\\\\
The only consistent interpretation of these circumstances would appear to be that the photometrically detectable  material in the region $R>R_0$ is detectable because it is excited to radiate above the background by the gravitational action of the core object, $\mathcal{M}_g(R \leq R_0)$. In old-fashioned terms, the core works the disk, thereby heating it. The  scaling relations then simply describe the effect of that action upon the disk and hence determine the mass of excited material and its radial distribution.
\subsection{$M_{flat}$\,: Theory compared to SPARC photometry}\label{Mflat}
The quantity $M_{flat}$ is conventionally defined as the mass contained within the disk out to $V_{rot} \approx V_{flat}$, and is estimated by integrating disk-photometry; consequently, by definition, $M_{flat}$ quantifies the detectable \emph{luminous} mass. However, in \S\ref{BTFR}, we found that the key to deriving the explicit BTFR, given at (\ref{eqn5c}) as
\begin{equation}
V_{flat}^4 = a_0\, G\, M_{flat}{\rm(theory)}, \label{eqn5b} 
\end{equation}
was the definition of (\ref{eqn5d}) that: 
\begin{equation}
M_{flat}{\rm(theory)} \equiv M_0\,\left(\frac{V_{flat}}{V_0} \right)^2.  \label{eqn5e}
\end{equation}
So the initial question to be answered is whether this definition  tracks $M_{flat}$(photometric) determined, for the SPARC sample, by  \citet{McGaugh2015}? 
\\\\
For this exercise,  we are testing $M_{flat}$(theory) directly against the specific $M_{flat}$(photometry) values provided by authors of the SPARC sample, and so take no account at all of the few special cases for which $V_0> V_{flat}$, discussed in \S\ref{SpecialCase-1}.
\subsubsection{The study of \citet{McGaugh2015}}
The work of \citet{McGaugh2015} was motivated by the idea that, within $\Lambda$CDM cosmology, the BTFR can only emerge from a complex process of galaxy formation, and is hence expected to be associated with significant intrinsic scatter. In short, the degree to which intrinsic scatter is present within the BTFR provides a key test for $\Lambda$CDM cosmology. The very high-quality of the SPARC sample provided an ideal opportunity to investigate BTFR scatter in a sample of substantial size. Subsequently, the authors were able to show that the SPARC sample is highly constrained by the BTFR showing far less scatter that expected from the $\Lambda$CDM model.
\\\\
But their results, in demonstrating a very tight fit of the empirical BTFR to SPARC data, also provide a test of  neo-MOND: specifically, to test whether or not $M_{flat}$(theory)  defined according to the scaling relation (\ref{eqn5e}) follows the photometric determinations of the same quantity given by \citet{McGaugh2015} over a sample of 118 SPARC objects.
\\\\
Our basic sample consists of the 129 SPARC non-bulgy objects with quality flag Q = 1 or 2. By contrast, the relevant sample of  \citet{McGaugh2015} consists of the 118 SPARC objects for which good estimations of $V_{flat}$ (using the standard photometric algorithm) were possible. Since some of these are bulgy objects (excluded from our sample), this gave a final sample of 85 SPARC objects of which to test $M_{flat}$(theory), given by (\ref{eqn5e}), against $M_{flat}$(photometry) given by  \citet{McGaugh2015}. 
\\\\
The results are displayed in the two panels of Figure\,\ref{fig:SSM-2} and discussed below.
\subsubsection*{Fig \ref{fig:SSM-2} Upper: Distance scales from conventional photometry}
\begin{figure}[H]
	\centering
	\includegraphics[width=0.65\linewidth]{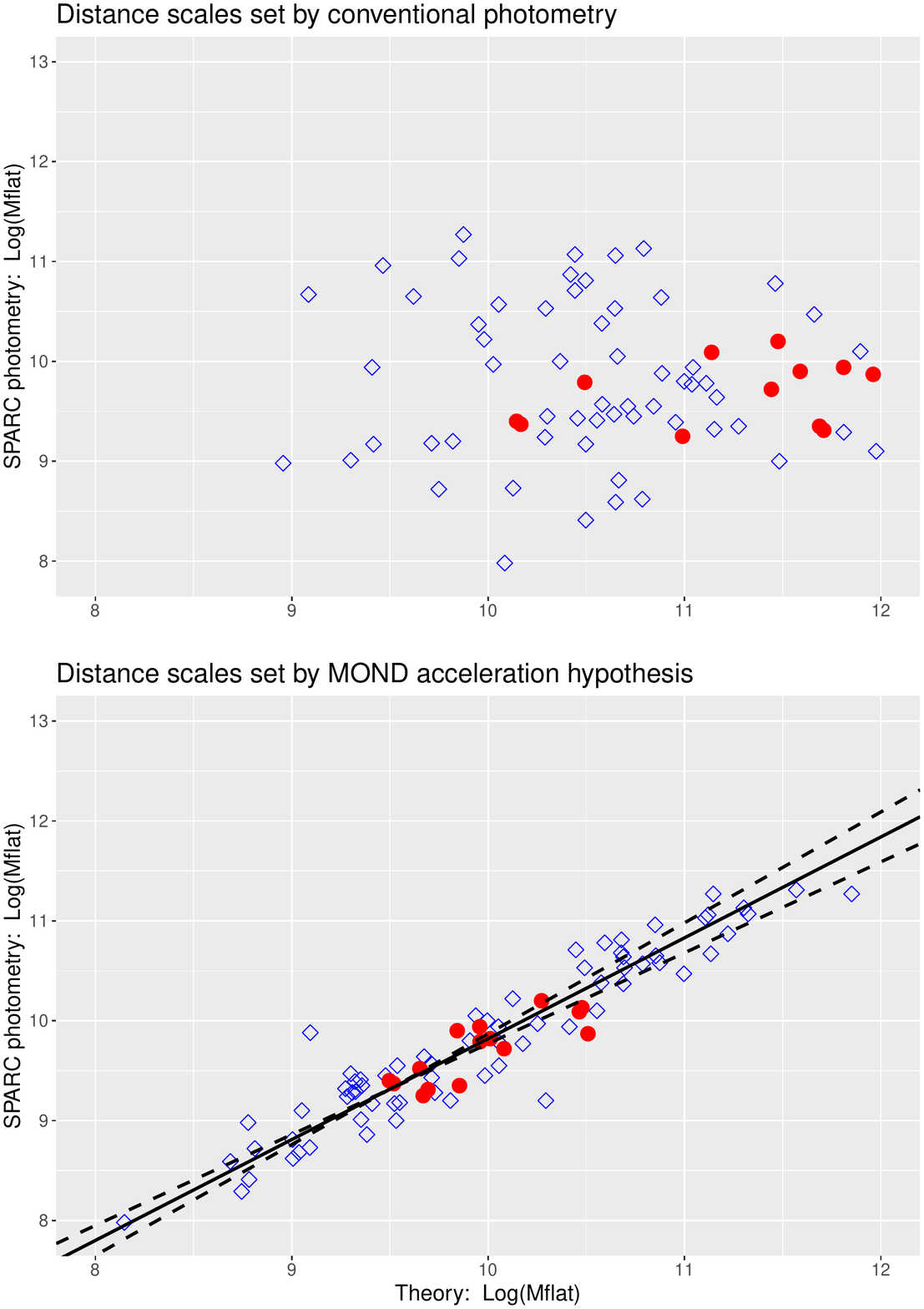}
	\caption{ Red filled circles = putative LSB objects. Open diamonds = everything else. Upper figure:  Here, the $\log{M_{flat}}$(theory) values have been computed using (\ref{eqn5e}) on the conventional photometric scalings implicit to the SPARC sample.  Lower figure: Here, the upper panel $\log{M_{flat}}$(theory) values have been rescaled according to the requirements of the MOND acceleration hypothesis.  No outliers have been removed.}
	\label{fig:SSM-2}
\end{figure}

Here, putative LSB objects are identified as red filled circles and $M_{flat}$(theory)  is computed on the photometric distance scalings implicit to the SPARC sample.
It is clear that almost all the plotted points lie in the region $M{\rm(photometry)} < M{\rm(theory)}$, again revealing a systematic \lq{missing mass}' mass problem.
\subsubsection*{Fig \ref{fig:SSM-2} Lower:  Distance scales from MOND acceleration hypothesis}
Here,  putative LSB objects are identified as red filled circles and $M_{flat}$(theory) computed above has been rescaled according to requirements of the MOND acceleration hypothesis, specified in \S\ref{Details}. 
The scatter plot of $\log{M_{flat}}({\rm photometry})$ against $\log M_{flat}$(theory) for this case makes it clear that this rescaling process  has the effect of mapping every data point almost perfectly onto the $\log M_{flat}$(theory) $=\log M_{flat}$(photometry) line. That is, there is now an almost statistically perfect correspondence between the two quantities and a least-area linear regression gives:
\begin{equation}
\log M_{flat}({\rm photometry}) =   \left(1.01 \pm 0.09 \right) \log M_{flat}({\rm theory}) -  \left(0.28 \pm 0.96 \right).  \label{eqn6d}
\end{equation}
Again, the imposition of the conditions required by the MOND acceleration hypothesis has the effect of imposing an almost perfect correspondence  $M_{flat}{\rm(theory)} \approx M_{flat}{\rm(photometry)}$, so that for all practical purposes, $M_{flat}$(theory) determined by (\ref{eqn5e}) tracks the \citet{McGaugh2015} photometric determinations of $M_{flat}$ with high statistical fidelity. The \lq{missing mass}' problem of the upper panel has disappeared completely.
\\\\
Finally, for completeness, a least-area regression between $\log M_{flat}$(photometry) and $\log V_{flat}$(theory) gives:
\begin{equation}
\log M_{flat}({\rm photometry}) =   \left(4.04 \pm 0.38 \right) \log V_{flat}({\rm theory}) +  \left(1.53 \pm 0.78 \right)  \label{eqn6dd}
\end{equation}
so that the BTFR is again confirmed on the SPARC data. A comparison between (\ref{eqn6dd}) and the corresponding regressions in 
\citet{McGaugh2015} shows that they are statistically identical, including the zero points. The only difference is that we are using $V_{flat}$(theory) rather than $V_{flat}$(SPARC).
\subsection{The radial distribution of luminous mass on the exterior: $R > R_0$} \label{Exterior}
Having considered $M_{flat}$(theory) against $M_{flat}$(photometry) on the SPARC sample, we are now in a position to model the totality of the radial distribution of luminous matter, $M_{Lum}(R)\leq M_{flat}$, on $R>R_0$. Whilst, strictly speaking, neo-MOND is silent on this topic, it does provide the scaling law
\begin{equation}
M_{flat}{\rm(theory)} \equiv M_0\,\left(\frac{V_{flat}}{V_0} \right)^2, \label{eqn8a} 
\end{equation}
which has been verified in \S\ref{Mflat},
as an indication of the possibilities.  Since we know from \S\ref{SpecialCase-1} that the cases $V_0 \geq V_{flat}$ imply that (\ref{eqn8a}) gives the mass contained within $R=R_{flat} \leq R_0$, then we need only consider the case $V_0< V_{flat}$ for which (\ref{eqn8a}) gives the luminous mass contained within $R > R_0$ as $V_{rot}(R)\rightarrow V_{flat}$. 
\\\\
This case is the most frequently occurring for the objects of the SPARC sample and, according to the refined Freeman's Law of (\ref{eqn4f}), is associated with the mass surface density condition $\Sigma_0 > \Sigma_F$ which, as is clear from \S\ref{RCbehaviour-1}, has a rotation curve characterized on $R>R_0$ by a smooth rise from $V_0$ to an asymptotic $V_{flat}$.
This suggests the possibility that a generalization of (\ref{eqn8a}) given by
\begin{equation}
M_{Lum}(R) = M_0 \left( \frac{V_{rot}(R)}{V_0} \right)^2,~~~R > R_0,~~~ V_{rot}(R) \leq V_{flat} \label{eqn7}
\end{equation}
will provide a means of calculating the radial distribution of luminous mass on $R > R_0$ for the $V_0 < V_{flat}$ object class. Alternatively, using (\ref{eqn2}b), this latter equation can be written:
\begin{equation}
M_{Lum}(R) = M_0 \left( \frac{V_{flat}}{V_0}\right)^2 \left(\frac{ \Sigma_F  R^2}{  \left(\Sigma_0 - \Sigma_F\right)R_0^2 + \Sigma_F R^2 } \right) ,~~ R > R_0 \label{eqn7a}
\end{equation}
so that the relationship between $M_{Lum}(R)$ and IGM is explicitly recognized.
\\\\
In order to test this generalized scaling law, we have integrated the photometry for each disk in the SPARC sample to obtain estimates of $M_{Lum}(R_i)(\rm{photometry})$ contained within radius $R_i$, for each radial coordinate $R_i > R_0$ on the measured disk; this gives a total of 1644 individual estimated mass values over the whole non-bulgy sample. The results for this test of the generalized scaling law (\ref{eqn7}) or (\ref{eqn7a}) are shown in figure \ref{fig:SSM-5}.
\subsubsection*{Fig \ref{fig:SSM-5} Upper: Distance scales from conventional photometry}
\begin{figure}[H]
	\centering
	\includegraphics[width=0.7\linewidth]{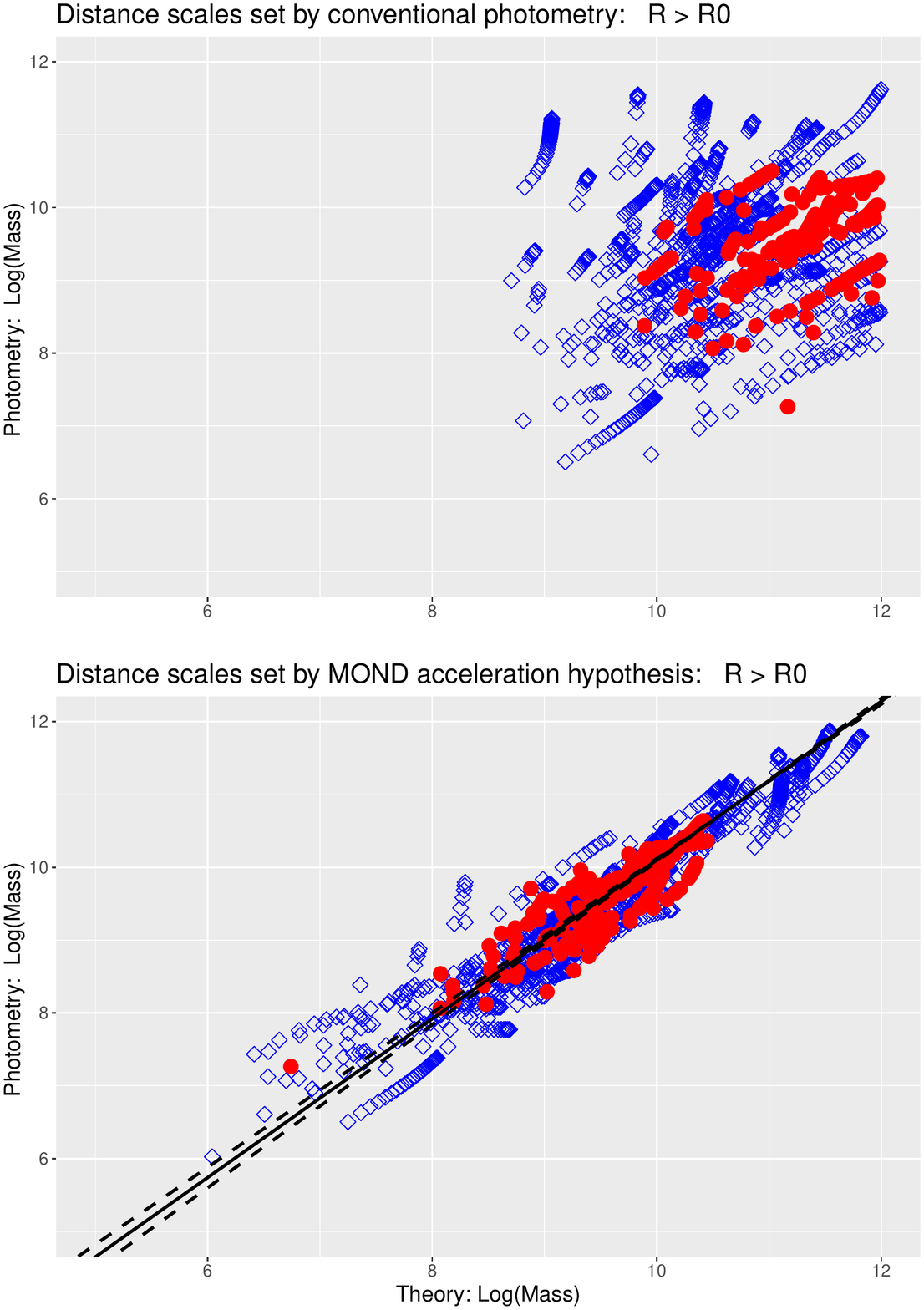}
	\caption{ Red filled circles = putative LSB objects. Open diamonds = everything else. Upper figure:  Here, the  $\log{M}(R)$(theory) values for all disk data points satisfying  $R > R_0$ (1644 individual points in total) have been computed using the conventional photometric distance scalings implicit to the SPARC sample.  Lower figure: Here, the 1644 $\log{M(R)}$(theory) values of the upper panel  have been rescaled according to the requirements of the MOND acceleration hypothesis.}
	\label{fig:SSM-5}
\end{figure}

Here, putative LSB objects are identified as red filled circles and $M$(theory) is computed by the algorithm of \S\ref{SA} on the photometric distance scalings implicit to the SPARC sample. The  scatter plot for this case makes it clear that for the most massive objects, say $10 \leq M \leq 12$, there is qualitative agreement between theory and photometry. However, at the low photometric end, we see that $M{\rm(theory)}> M{\rm(photometry)}$ by up to two orders of magnitude. That is, a \lq{missing mass}' problem of similar proportions to that recognized in classical theory has emerged once more.
\subsubsection*{Fig \ref{fig:SSM-5} Lower: Distance scales from MOND acceleration hypothesis}
Here, putative LSB objects are identified as red filled circles and $M$(theory) computed above have been rescaled according to the requirements of the MOND acceleration hypothesis (\ref{eqn3NN}a) which are specified in \S\ref{Details}.
The  scatter plot for this case makes it clear that, in broad terms, the rescaling process has the bulk effect of shifting $\log{M}({\rm photometry}) < 10.0$ objects to the left in the plot to give an obvious \& powerful correlation. A least-area linear regression gives:
\begin{eqnarray}
\log{M}({\rm photometry}) &\approx& \left(1.09 \pm 0.03\right) \log{M}({\rm theory}) - (0.80 \pm 0.32),
\nonumber 
\end{eqnarray}
som that $\log M(\rm{theory}) \approx \log M(\rm{photometry})$ is confirmed for all the mass measurements outside the critical radius, $R > R_0$.
\\\\
We can unambiguously conclude that the scaling relation (\ref{eqn7}) (or equivalently (\ref{eqn7a})) is validated on the SPARC sample.
\section{Qualitative behaviour of RCs on $R>R_0$} \label{RCbehaviour-1}
It is easily shown from (\ref{eqn2}) that, for $R>R_0$: 
\[
\frac{d V_{rot}}{dR} \sim \frac{k^2}{R^3} \left( \Sigma_0 - \Sigma_F \right),~~~k^2>0,~~~R>R_0.
\]
As a direct consequence, we can then deduce:
\begin{itemize}
	\item if $\Sigma_0 > \Sigma_F$, then rotation curves on $R\leq R_0$ reach $(R_0,V_0)$ and then rise smoothly with an asymptotic approach to a constant rotation velocity $V_{flat} > V_0$; 
	\item if $\Sigma_0 = \Sigma_F$, then rotation curves on $R \leq R_0$ reach $(R_0,V_0)$ and then change abruptly to flatness with constant rotation velocity $V_{flat} = V_0$;
	\item if $\Sigma_0 < \Sigma_F$, then rotation curves on $R\leq R_0$ reach $(R_0,V_0)$ and then fall smoothly with an asymptotic approach to a constant velocity $V_{flat}< V_0$. 
\end{itemize}
\section{Summary and conclusions}
 \subsection{Summary}
The foregoing for all objects can be entirely summarized by reviewing the case of LSB galaxies.
\\\\ 
Neo-MOND is so-called because it shares with MOND Milgrom's fundamental idea of the critical acceleration radius at which gravitational acceleration reaches the critical value of $a_0 \approx 1.2 \times 10^{-10} m/sec^2$.  For both MOND and neo-MOND, this critical acceleration radius defines the position at which one set of physical circumstances gives way to another. 
\\\\
In each case,  the transition between one set of physical circumstances and another creates a radial \emph{gradient} discontinuity in the rotation curve. MOND considers this discontinuity to be an artifact of its procedures (and eliminates it using an interpolation process) whilst neo-MOND considers it to be an objective \& fundamental feature of rotation curves in general. 
\\\\
Given that such discontinuities can be directly \& reliably detected, it follows immediately that the MOND acceleration hypothesis of \S\ref{MOND-AC} provides the means of setting distance scales which is completely independent of standard candles and the photometric method:  we simply adjust the distance to the galaxy concerned (and hence all associated linear scales)  until the centripetal acceleration (used as a proxy for gravitational acceleration) at the detected gradient discontinuity satisfies $V^2/R \approx a_0$.
\\\\
Now consider the case of low surface brightness disk galaxies: LSB disks are defined as disks for which $V^2/R < a_0$ everywhere, and it is considered to be one of MOND's great successes that such apparently weakly gravitating objects were predicted to exist before they were first observed in the early 1990s. It follows that LSB disks, by definition, do not possess the radial gradient discontinuity  in their rotation curves, and therefore the MOND acceleration hypothesis cannot be used to set a distance scale for them.
\\\\
Now  consider the SPARC sample: according to the notes in the source papers used by \citet{McGaugh2015} to compile that sample, it contains at least 25 LSB objects, listed in \S\ref{LSBs}. But figures \ref{fig:SPARCMASSvsTheoryMass}, 	\ref{fig:SSM-4}, \ref{fig:SSM-2} \&  \ref{fig:SSM-5} demonstrate beyond all doubt that these putative LSB objects behave exactly as do all the other objects in the SPARC sample under the MOND acceleration hypothesis. In other words, as a matter of objective fact, they must necessarily all possess a radial gradient discontinuity in their rotation curves, which has been successfully located by the algorithm of \S\ref{Details}, and then automatically acted upon via the requirements of the acceleration hypothesis, with the final result of bringing their masses exactly (in a statistical sense) onto the line $Mass{\rm(theory)}=Mass{\rm(photometric)}$ and thereby dissolving the missing mass problem.
\\\\
There is only one way this objective fact can be made consistent with the appearance that $V^2/R < a_0$ over the resolvable disk for all the objects classified as LSBs: there must be some unrecognized mechanism which systematically exaggerates distance scales, and hence the linear scales of these objects, when these are determined photometrically, giving rise to the appearance that $V^2/R < a_0$ everywhere over the resolvable disk.
\\\\
Since figures \ref{fig:SPARCMASSvsTheoryMass}, 	\ref{fig:SSM-4}, \ref{fig:SSM-2} \&  \ref{fig:SSM-5} show that what is true for the putative LSB objects in the sample is also true for all objects in the sample, then the unrecognized mechanism which exaggerates the photometric distance scales of the putative LSBs does the same for all the objects in the SPARC sample.  The one obvious possibility available to us, and already discussed in \S\ref{GeneralConsiderations},  \S\ref{Distances1} and \S\ref{Distances2}, is that we are detecting indirect evidence for the  $D\approx 2$ quasi-fractal grey-dust IGM which is a necessary component of the neo-MOND worldview.
\subsection{Conclusions}
In the following, the phrase \lq{luminous mass}' simply refers to that material which is photometrically detectable and so necessarily radiates above the background of which the material component is, according to neo-MOND, the $D\approx 2$ fractal grey-dust IGM. 
\\\\
With this understanding, the implication of the foregoing analysis of the SPARC sample is that the conventional model of a disk galaxy, considered to consist of a core of luminous material, surrounded by luminous gas and set into a very large gravitationally bound sphere of undetected material of an unknown quality, is replaced by a model considered to consist of a core of luminous material set into the unbounded $D\approx 2$ quasi-fractal grey-dust IGM, assumed to have a luminous gaseous component in the $R>R_0$ local galactic environment, with which it interacts in a manner orchestrated by the primary neo-MOND scaling relations
\begin{eqnarray}
\frac{V_{rot}(R)}{V_{flat}} &=&  \left(\frac{ \Sigma_F }{   \Sigma_R } \right)^{1/2}, ~~~~ R \leq R_0; \nonumber \\
\nonumber \\
\frac{V_{rot}(R)}{V_{flat}} &=&  \left(\frac{ \Sigma_F  R^2}{  \left(\Sigma_0 - \Sigma_F\right)R_0^2 + \Sigma_F R^2 } \right)^{1/2},~~~~ R > R_0, \nonumber \\
\nonumber 
\end{eqnarray}
together with the secondary scaling relation (\ref{eqn7}) inferred as a generalization from the scaling relation (\ref{eqn8a}) and rearranged for consistency with the primary scaling relations above as:
\[
\frac{V_{rot}(R)}{V_{flat}} = \frac{V_0}{V_{flat}}\left(\frac{ \Sigma_R\,R^2}{\Sigma_0\,R_0^2} \right)^{1/2},~~~R > R_0,~~~ V_{rot}(R) \leq V_{flat}.
\]
Within these scaling relations,   $\Sigma_R$ is the mass surface density of luminous mass at radius $R$, $\Sigma_0$ is the mass surface density of luminous mass at $R_0$, $\Sigma_F\equiv a_0/(4\pi G)$ is interpreted to be the characteristic mass surface density of the $D\approx 2$ quasi-fractal grey dust IGM and the radial gradient discontinuity at  $R_0$ is considered to be an objective feature of the rotation curve which is conjectured to coincide with the place at which the gravitational acceleration reaches the MOND limit, $a_0 \approx 1.2\times 10^{-10}\,\rm{mtrs/sec}^2$.
\\\\
This latter conjecture is \emph{the MOND acceleration hypothesis} which is fundamental to the analysis presented here and  provides a means of setting distance scales which is entirely independent of standard candles and the photometric method.
\appendix
\section{Outline derivation of neo-MOND}\label{Outline}
The disk galaxy scaling relationships provided by neo-MOND arise as a very special case solution of a general theory based upon the ideas of Liebniz and Mach, which is developed in detail in the arXiv paper \citet{Roscoe2018}. There are three inputs to neo-MOND: firstly, that which is specific to the underlying theory; secondly, that which is specific to Milgrom's MOND phenomenology; thirdly, the MOND acceleration hypothesis.
\subsection{The underlying theory content: the equilibrium case} \label{Equilibrium world}
The most simple case of the general theory is that of a world in global equilibrium and, according to the general theory, material in this equilibrium world is distributed fractally, with fractal dimension $D=2$ so that, \emph{about any origin}, material is distributed according to
\[
\mathcal{M}_F(R) \equiv 4 \pi \Sigma_F R^2
\]  
where $\Sigma_F$ is the mass surface density, which is a global constant in a fractal $D=2$ world. This most simple case of the general theory conforms exactly with the quasi-fractal $D\approx 2$ distribution of material on medium cosmological scales supported by the discussion of \S\ref{Observations}, and argued for by  \citet{Baryshev1995} and others over many years.
\\\\
This latter simple case represents our starting point: to construct neo-MOND on the basis of it, we begin by assuming the existence of a finite, but otherwise unspecified, spherically symmetric mass perturbation, $\mathcal{M}(R)$ say, of the equilibrium fractal environment, $\mathcal{M}_F (R)$. This perturbation creates a specific centre and, by definition, for such a system,
\begin{equation}
\mathcal{M}(R) \rightarrow \mathcal{M}_F(R)\equiv 4\pi \Sigma_F R^2 ~~~\rm{as}~~~R \rightarrow \infty.
\label{eqn4gg}
\end{equation}
According to the general theory (\citet{Roscoe2018}), the dynamics associated with an arbitrary $\mathcal{M}(R)$ admit a degenerate state in which only circular motions can occur, and for which the associated energy equation is given by
\begin{equation}
V^2_{rot}(R) \mathcal{M}(R)  - V^2_{flat} 4 \pi \Sigma_F R^2 = 0. \label{eqn4g}
\end{equation}
This rearranges as
\begin{equation}
V_{rot}(R) = V_{flat} \, \left( \frac{4 \pi \Sigma_F R^2}{\mathcal{M}(R)} \right)^{1/2} \label{eqn4}
\end{equation}
from which, using (\ref{eqn4gg}), we see that $V_{rot} \rightarrow V_{flat}$ as $R \rightarrow \infty$ so that $V_{flat}$ is an asymptotic flat rotation velocity. 
\\\\
Now, whilst a disk is, by definition, not spherical, it does sit within its external environment which is spherical. So, our very simple model assumes spherical symmetry, and that all motions take place within the equatorial plane, with the rearranged energy equation (\ref{eqn4}) providing the theoretical foundation of neo-MOND.
\subsection{The MOND phenomenological content} 
We begin by noting the fact that the existence of the  mass surface density scale, $\Sigma_F$, in a fractal $D=2$ world implies the existence of a corresponding acceleration scale, $a_F$, the two being connected via the gravitational constant:
\[
a_F = 4 \pi \Sigma_F\,G. 
\] 
If we now consider the evidence of Milgrom's MOND that, within a galaxy disk, there is a critical gravitational acceleration scale, $a_0$ say, with a corresponding critical radius, $R_0$, at which the critical gravitational acceleration is reached, then it becomes natural to make the association $a_0 \equiv a_F$ so that
\[
a_0 = 4 \pi \Sigma_F\,G \label{eqn4c}
\]
and then to hypothesize that $R_0$ defines the boundary between the interior environment of a disk galaxy and a transition zone which will ultimately merge into the exterior equilibrium environment. This, in effect, is the MOND acceleration hypothesis of \S\ref{MOND-AC} On the basis of this hypothesis, and in the specific context of modelling a galaxy, we can deduce that $\mathcal{M}(R)$ in (\ref{eqn4}) must have the general structure:
\begin{eqnarray}
\mathcal{M}(R) &=& \mathcal{M}_g (R), \,\,\, R \leq R_0 \nonumber \\
\label{eqn4h} \\
\mathcal{M}(R) &=&  \mathcal{M}_g(R_0) + 4 \pi \Sigma_F \left(R^2 - R^2_0 \right),\,\,\, R > R_0 \nonumber
\end{eqnarray}
where 
\begin{itemize}
\item $\mathcal{M}_g(R)$ is the model for the mass distribution  within the galaxy interior, $R\leq R_0$;
\item $\mathcal{M}_g(R_0)$  is the total mass (consisting of stars, dust and gas) contained within the critical acceleration boundary, $R \leq R_0$;
\item since the material on the exterior of the critical acceleration boundary is primarily composed of gas/dust, then the component $ 4 \pi \Sigma_F \left(R^2 - R^2_0 \right)$ in any annular region $R > R_0$ consists, by definition, primarily of gas/dust.
\end{itemize}
The system's energy equation rearranged as (\ref{eqn4}) and combined  (\ref{eqn4h})  gives the general case neo-MOND model:
\begin{eqnarray}
V_{rot}(R) &=& V_{flat} \left(\frac{ 4 \pi \Sigma_F  R^2}{ \mathcal{M}_{g}(R) } \right)^{1/2}, ~~~~0< R \leq R_0; \nonumber \\
V_{rot}(R) &=& V_{flat} \left(\frac{ 4 \pi \Sigma_F  R^2}{ \mathcal{M}_g(R_0)  + 4 \pi \Sigma_F (R^2-R_0^2) } \right)^{1/2},~~~~ R > R_0 \nonumber \\
\nonumber
\end{eqnarray}
which can be conveniently rearranged as:
\begin{eqnarray}
V_{rot}(R) &=& V_{flat} \left(\frac{ \Sigma_F }{   \Sigma_R } \right)^{1/2}, ~~~~ R \leq R_0; \nonumber \\
V_{rot}(R) &=& V_{flat} \left(\frac{ \Sigma_F  R^2}{  \left(\Sigma_0 - \Sigma_F\right)R_0^2 + \Sigma_F R^2 } \right)^{1/2},~~~~ R > R_0, \label{eqn4a} 
\end{eqnarray}
where
\[
\Sigma_R \equiv \frac{\mathcal{M}_g(R)}{4 \pi \,R^2},~~~\Sigma_0 \equiv \frac{\mathcal{M}_g(R_0)}{4 \pi\,R_0^2}.
\] 
\section{Some computational details}\label{MassModels}
There are various details which are necessary to reliably reproduce the results of this paper.
\subsection*{The minimization metric}
For minimization problems involving noisy data, it is generally considered best to use a metric based on the $L_1$-norm. So, for a rotation curve with measured velocities $V_{sparc}(R_i),\,i=1..N$, we seek to determine the disk parameters $(R_0, \Sigma_0, V_{flat})$ of  (\ref{eqn4a}) by minimizing:
\[
metric = \sum^N_{i=1} \left| \frac{V_{sparc}(R_i)-V_{theory}(R_i)}{V_{sparc}(R_i)} \right|
\]
with respect to variation in them. This gives far superior results to those arising from use of the $L_2$-norm, for example.
\subsection*{The Nelder-Mead iteration}
Because the data is noisy, it is necessary to run the Nelder-Mead minimization process multiple times for each disk, with randomly generated initial guesses for the RC fitting parameters  $(R_0, \Sigma_0, V_{flat})$. In practice, this means running the minimization process (typically) about 2000 times per disk before the results completely settle down.
\section{Least-areas linear regression} \label{LeastAreas}
\begin{figure}[H]
	\centering
	\includegraphics[width=0.6\linewidth]{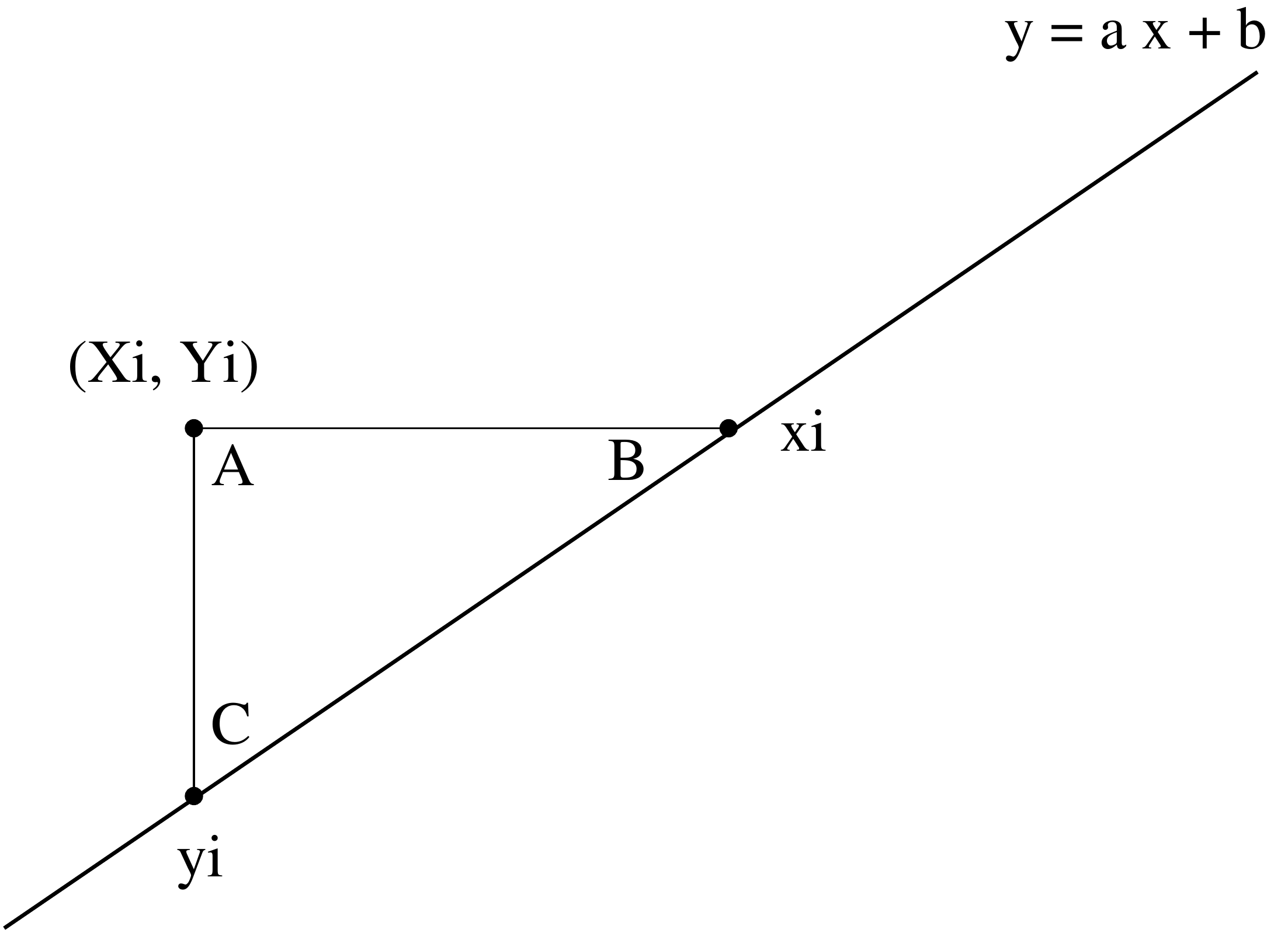}
	\label{Fig:LeastAreas1}
\end{figure}

Suppose that we have the data $(X_i, Y_i),~i = 1..N$ to which we fit the model
\begin{equation}
y = A_0 +A_1 x  \label{LA1}
\end{equation}
according to a criterion labelled as \emph{least areas} which we  describe below.
\\\\
From figure \ref{Fig:LeastAreas1}, the area of the triangle shown is given by
\begin{equation}
\Delta_i = -\frac{1}{2}\left(X_i - x_i \right) \left(Y_i - y_i \right), \label{LA2}
\end{equation}
which is always positive, regardless of the position of the point $A$.
From (\ref{LA1}):
\[
x_i =  - \frac{A_0}{A_1} + \frac{1}{A_1 }Y_i,~ ~~~y_i = A_0 +  A_1 X_i 
\]
so that (\ref{LA2}) becomes:
\[
\Delta_i = -\frac{1}{2}\left(X_i  + \frac{A_0}{A_1} - \frac{1}{A_1 }Y_i  \right) \left(Y_i -A_0 - A_1 X_i \right).
\]
Dropping the numerical factor, the least-area regression model is derived by minimizing
\[
Area = \sum_{i=1}^N \Delta_i = \sum_{i=1}^N \left(X_i  + \frac{A_0}{A_1} - \frac{1}{A_1 }Y_i  \right) \left(Y_i -A_0 - A_1 X_i \right)
\] 
with respect to variations in $A_0$ and $A_1$. The normal equations are found to be:
\begin{eqnarray}
N A_0 +  A_1 \Sigma X_i  &=& \Sigma Y_i \nonumber \\
\nonumber \\
N A_0^2 -2 A_0  \Sigma Y_i - A_1^2
 \Sigma X_i^2 &=& - \Sigma Y_i^2.
\end{eqnarray}
Hence: 
\[
A_1 = \pm \sqrt{\frac{\left(\Sigma Y_i\right)^2-N \Sigma Y_i^2}{\left(\Sigma X_i\right)^2-N \Sigma X_i^2} }, ~~~
A_0 = \frac{\left( \Sigma Y_i - A_1 \Sigma X_i\right)}{N}
\]
for the result.


\newpage
\begin{thebibliography}{99}
\bibitem[\protect\citeauthoryear{Aguirre}{1999}]{AA1999}	Aguirre, A., 1999, ApJ, 525, 583
\bibitem[\protect\citeauthoryear{Baryshev et al}{1995}]{Baryshev1995} Baryshev, Yu V., Sylos Labini, F., Montuori, M., Pietronero, L. 1995 \textit{Vistas in Astronomy} 38, 419
\bibitem[\protect\citeauthoryear{Broadhurst et al}{1990}]{Broadhurst}Broadhurst, T.J., Ellis, R.S., Koo, D.C., Szalay, A.S., 1990, Nat 343, 726
\bibitem[\protect\citeauthoryear{Charlier}{1908}]{Charlier1908}Charlier, C.V.L., 1908, Astronomi och Fysik 4,
1 (Stockholm)
\bibitem[\protect\citeauthoryear{Charlier}{1922}]{Charlier1922}Charlier, C.V.L., 1922, Ark. Mat. Astron. Physik
16, 1 (Stockholm)
\bibitem[\protect\citeauthoryear{Charlier}{1924}]{Charlier1924}Charlier, C.V.L., 1924, Proc. Astron. Soc. Pac. 37, 177

\bibitem[\protect\citeauthoryear{Corasaniti}{2006}]{PSC2006}Corasaniti, P.S., 2006, MNRAS 372, 1, 191-198 



\bibitem[\protect\citeauthoryear{Courteau}{1997}]{SC1997} Courteau S., 1997, AJ, 114, 6, 2402-2427


\bibitem[\protect\citeauthoryear{Da Costa et al}{1994}]{DaCosta} Da Costa, L.N., Geller, M.J., Pellegrini, P.S., Latham, D.W., Fairall, A.P., Marzke, R.O., Willmer, C.N.A., Huchra,
J.P., Calderon, J.H., Ramella, M., Kurtz, M.J., 1994, ApJ 424, L1

 

\bibitem[\protect\citeauthoryear{Dale, Giovanelli \& Haynes}{1997}]{Dale1997} Dale DA, Giovanelli R, Haynes M, 1997
AJ 114 (2): 455-473 

\bibitem[\protect\citeauthoryear{Dale et al}{1998}]{Dale1998} Dale DA, Giovanelli R, Haynes MP, Scodeggio M, Hardy E, Campusano LE, 1998, AJ 115 (2), 418-435


\bibitem[\protect\citeauthoryear{Dale, Giovanelli \& Haynes}{1999}]{Dale1999} Dale DA, Giovanelli R, Haynes MP,  1999, AJ 118 (4), 1468-1488

\bibitem[\protect\citeauthoryear{Dale \& Uson}{2000}]{Dale2000} Dale D.A., Uson JM, 2000, AJ 120 (2), 552-561


\bibitem[\protect\citeauthoryear{Dale et al}{2001}]{Dale2001}
Dale D.A., Giovanelli R, Haynes M.P., Hardy E, Campusano LE, 2001,
AJ 121, 1886-1892 




\bibitem[\protect\citeauthoryear{De Lapparent et al}{1988}]{DeLapparent1988} De Lapparent, V., Geller,M.J., Huchra, J.P., 1988, ApJ 332, 44
\bibitem[\protect\citeauthoryear{De Vaucouleurs}{1970}]{De Vaucouleurs1970} De Vaucouleurs, G., 1970, Sci 167, 1203
\bibitem[\protect\citeauthoryear{Gabrielli \& Sylos Labini}{2001}]{Gabrielli} Gabrielli, A., Sylos Labini, F., 2001, Europhys.
Lett. 54 (3), 286
\bibitem[\protect\citeauthoryear{Giovanelli and Haynes}{1986}]{Giovanelli1986} Giovanelli, R., Haynes, M.P., Chincarini, G.L., 1986, ApJ 300, 77
\bibitem[\protect\citeauthoryear{Huchra et al}{1983}]{Huchra1983} Huchra, J., Davis, M., Latham, D.,Tonry, J., 1983, ApJS 52, 89
\bibitem[\protect\citeauthoryear{Hogg et al}{2005}]{Hogg} Hogg, D.W., Eistenstein, D.J., Blanton M.R., Bahcall N.A, Brinkmann, J., Gunn J.E., Schneider D.P. 2005 ApJ, 624, 54

\bibitem[\protect\citeauthoryear{Joyce, Montuori \& Sylos Labini et al}{1999}]{Joyce} Joyce, M., Montuori, M., Sylos Labini, F., 1999,
Astrophys. J. 514, L5
\bibitem[\protect\citeauthoryear{Labini \& Gabrielli}{2000}]{Labini} Labini, F.S., Gabrielli, A., 2000, \emph{Scaling and fluctuations in galaxy distribution: two tests to probe large scale structures}, astro-ph0008047
\bibitem[\protect\citeauthoryear{Lelli, McGaugh \& Schombert}{2015}]{McGaugh2015} Lelli, F., McGaugh, SS, Schombert, JM., arxiv.org/abs/1512.04543
\bibitem[\protect\citeauthoryear{Martinez et al}{1998}]{Martinez} Martinez, V.J., PonsBorderia, M.J., Moyeed, R.A., Graham, M.J. 1998 \textit{MNRAS} 298, 1212 
\bibitem[\protect\citeauthoryear{Mathewson, Ford \& Buchhorn}{1992}]{MFB1992} Mathewson, D.S., Ford, V.L., Buchhorn, M., 1992, {Astrophys J. Supp.} {81}, 413

\bibitem[\protect\citeauthoryear{Mathewson \& Ford}{1996}]{MF1996} Mathewson, D.S., Ford, V.L., 1996, {Astrophys J. Supp.} {107}, 97

\bibitem[\protect\citeauthoryear{McGaugh}{1995a}]{McGaugh1995a} McGaugh SS, Schombert JM, Bothun GD. 1995a. Astron. J. 109: 2019-2033
\bibitem[\protect\citeauthoryear{McGaugh}{1995b}]{McGaugh1995b} McGaugh SS, Bothun GD, Shombert JM. 1995b. Astron. J. 110: 573-580
\bibitem[\protect\citeauthoryear{McGaugh}{1996}]{McGaugh1996} McGaugh, SS. 1996. MNRAS 280: 337-354
\bibitem[\protect\citeauthoryear{McGaugh}{1998a}]{McGaugh1998a} McGaugh, SS. 1998a. In After the Dark Ages: When Galaxies Were Young, eds. Holt, S.S. \& Smith, E.P.
pp 72-75. AIP (astro-ph/9812328)
\bibitem[\protect\citeauthoryear{McGaugh}{1998b}]{McGaugh1998b} McGaugh, SS, de Blok, WJG. 1998b. Ap. J. 499: 41-65
\bibitem[\protect\citeauthoryear{McGaugh}{1998c}]{McGaugh1998c} McGaugh, SS, de Blok, WJG. 1998c. Ap. J. 499: 66-81
\bibitem[\protect\citeauthoryear{McGaugh}{1999a}]{McGaugh1999a} McGaugh, SS. 1999a. In Galaxy Dynamics, eds. Merritt, D., Sellwood, J.A., Valluri, M., pp. 528-538.
San Francisco: Astron.Soc.Pac (astro-ph/9812327)
\bibitem[\protect\citeauthoryear{McGaugh}{1999b}]{McGaugh1999b} McGaugh, SS. 1999b. Ap. J. Lett. 523: L99-L102
\bibitem[\protect\citeauthoryear{McGaugh}{2000a}]{McGaugh2000a} McGaugh, SS. 2000a. Ap. J. Lett. 541: L33-L36
\bibitem[\protect\citeauthoryear{McGaugh}{2000b}]{McGaugh2000b} McGaugh, SS, Schombert, JM, Bothun, GD, de Blok, WJG. 2000b. Ap. J.
\bibitem[\protect\citeauthoryear{McGaugh}{2001}]{McGaugh2001} McGaugh, SS, Rubin, VC, de Blok, WJG. 2001. Astron. J. 122: 2381-2395
\bibitem[\protect\citeauthoryear{Milgrom}{1983a}]{Milgrom1983a}	Milgrom, M., 1983a, Astrophysical Journal. 270: 365. 
\bibitem[\protect\citeauthoryear{Milgrom}{1983b}]{Milgrom1983b} Milgrom, M., 1983b, Astrophysical Journal. 270: 371
\bibitem[\protect\citeauthoryear{Milgrom}{1983c}]{Milgrom1983c} Milgrom, M. 1983c. Ap. J. 270: 365-370
\bibitem[\protect\citeauthoryear{Milgrom}{1983d}]{Milgrom1983d} Milgrom, M. 1983d. Ap. J. 270: 371-383
\bibitem[\protect\citeauthoryear{Milgrom}{1983e}]{Milgrom1983e} Milgrom, M. 1983e. Ap. J. 270: 384-389
\bibitem[\protect\citeauthoryear{Milgrom}{1984}]{Milgrom1984} Milgrom, M. 1984. Ap. J. 287: 571-576
\bibitem[\protect\citeauthoryear{Milgrom}{1988}]{Milgrom1988} Milgrom, M. 1988. Astron. Astrophys. 202: L9-L12
\bibitem[\protect\citeauthoryear{Milgrom}{1989a}]{Milgrom1989a} Milgrom, M. 1989a. Ap. J. 338: 121-127
\bibitem[\protect\citeauthoryear{Milgrom}{1989b}]{Milgrom1989b} Milgrom, M. 1989b. Astron. Astrophys. 211: 37-40
\bibitem[\protect\citeauthoryear{Milgrom}{1989c}]{Milgrom1989c} Milgrom, M. 1989c, Comments Astrophys. 13: 215-230
\bibitem[\protect\citeauthoryear{Milgrom}{1991}]{Milgrom1991} Milgrom, M. 1991. Ap. J. 367: 490-492
\bibitem[\protect\citeauthoryear{Milgrom}{1994a}]{Milgrom1994a} Milgrom, M. 1994a. AnnalsPhys 229: 384-415
\bibitem[\protect\citeauthoryear{Milgrom}{1994b}]{Milgrom1994b} Milgrom, M. 1994b. Ap. J. 429: 540-544
\bibitem[\protect\citeauthoryear{Milgrom}{1995}]{Milgrom1995} Milgrom, M. 1995. Ap. J. 455: 439-442
\bibitem[\protect\citeauthoryear{Milgrom}{1997a}]{Milgrom1997a} Milgrom, M. 1997a. Phys.Rev.E 56: 1148-1159
\bibitem[\protect\citeauthoryear{Milgrom}{1997b}]{Milgrom1997b} Milgrom, M. 1997b. Ap. J. 478: 7-12
\bibitem[\protect\citeauthoryear{Milgrom}{1998}]{Milgrom1998} Milgrom, M. 1998. Ap. J. Lett. 496. L89-L91
\bibitem[\protect\citeauthoryear{Milgrom}{1999}]{Milgrom1999} Milgrom, M. 1999. Phys. Lett. A 253: 273-279
\bibitem[\protect\citeauthoryear{Milgrom}{2002}]{Milgrom2002} Milgrom, M. 2002. Ap. J. Lett. 577: L75-L77 
\bibitem[\protect\citeauthoryear{Peebles}{1980}]{Peebles1980} Peebles, P.J.E., 1980, The Large Scale Structure of the Universe, Princeton University Press, Princeton, NJ.
\bibitem[\protect\citeauthoryear{Persic \& Salucci}{1995}]{PS1995}
Persic M., Salucci P., 1995, {ApJS}, 99, 501
\bibitem[\protect\citeauthoryear{Pietronero \& Sylos Labini}{2000}]{Pietronero} Pietronero, L., Sylos Labini, F., 2000, Physica
A, (280), 125

\bibitem[\protect\citeauthoryear{Roscoe}{1999}]{Roscoe1999} Roscoe D.F., 1999, {A\&A}, {343}, 788-800
\bibitem[\protect\citeauthoryear{Roscoe}{2002A}]{Roscoe2002A} Roscoe D.F., 2002A, {General Relativity \& Gravitation}, {34}, 577-603
\bibitem[\protect\citeauthoryear{Roscoe}{2002B}]{Roscoe2002B} Roscoe D.F., 2002B, {A\&A}, {385}, 431-453
\bibitem[\protect\citeauthoryear{Roscoe}{2004}]{Roscoe2004} Roscoe D.F., 2004, {General Relativity \& Gravitation}, {36}, 3-45
\bibitem[\protect\citeauthoryear{Roscoe}{2018}]{Roscoe2018} Roscoe D.F., A complete Leibniz-Mach cosmology:  arxiv.org/abs/0802.2889
\bibitem[\protect\citeauthoryear{Sanders}{1984}]{Sanders1984} Sanders, RH. 1984. Astron. Astrophys. 136: L21-L23
\bibitem[\protect\citeauthoryear{Sanders}{1986}]{Sanders1986} Sanders, RH. 1986. MNRAS 223: 539-555
\bibitem[\protect\citeauthoryear{Sanders}{1988}]{Sanders1988} Sanders, RH. 1988. MNRAS 235: 105-121
\bibitem[\protect\citeauthoryear{Sanders}{1989}]{Sanders1989} Sanders, RH. 1989. MNRAS 241: 135-151
\bibitem[\protect\citeauthoryear{Sanders}{1990}]{Sanders1990} Sanders, RH. 1990. Astron.Astrophys. Rev. 2: 1-28
\bibitem[\protect\citeauthoryear{Sanders}{1994a}]{Sanders1994a} Sanders, RH. 1994a. Astron. Astrophys. 284: L31-L34
\bibitem[\protect\citeauthoryear{Sanders}{1994b}]{Sanders1994b} Sanders, RH, Begeman, KG, 1994b, MNRAS , 266: 360-366
\bibitem[\protect\citeauthoryear{Sanders}{1996}]{Sanders1996} Sanders, RH. 1996. Ap. J. 473: 117-129
\bibitem[\protect\citeauthoryear{Sanders}{1997}]{Sanders1997} Sanders, RH. 1997. Ap. J. 480: 492-502
\bibitem[\protect\citeauthoryear{Sanders}{1998a}]{Sanders1998a} Sanders, RH. 1998a. MNRAS 296: 1009-1018
\bibitem[\protect\citeauthoryear{Sanders}{1998b}]{Sanders1998b} Sanders, RH, Verheijen MAW. 1998b Ap. J. , 503, 97-108
\bibitem[\protect\citeauthoryear{Sanders}{1999}]{Sanders1999} Sanders, RH. 1999. Ap. J. 512: L23-L26
\bibitem[\protect\citeauthoryear{Sanders}{2000}]{Sanders2000} Sanders, RH. 2000. MNRAS 313: 767-774
\bibitem[\protect\citeauthoryear{Sanders}{2002}]{Sanders2002} Sanders, R., https://arxiv.org/abs/astro-ph/0212293
\bibitem[\protect\citeauthoryear{Sanders \& McGaugh}{2002}]{Sanders2002A} Sanders, R., McGaugh, S., https://arxiv.org/abs/astro-ph/0204521v1
\bibitem[\protect\citeauthoryear{Sanders}{2001}]{Sanders2001} Sanders, RH. 2001. Ap. J. 560: 1-6
\bibitem[\protect\citeauthoryear{Sanders}{2014}]{Sanders2014} Sanders, R. H., 2014, Canadian Journal of Physics. 93 (2): 126
\bibitem[\protect\citeauthoryear{Scaramella et al}{1998}]{Scaramella} Scaramella, R., Guzzo, L., Zamorani, G., Zucca,
E., Balkowski, C., Blanchard, A., Cappi, A., Cayatte, V., Chincarini,
G., Collins, C., Fiorani, A., Maccagni, D., MacGillivray, H., Maurogordato,
S., Merighi, R., Mignoli, M., Proust, D., Ramella, M., Stirpe, G.M.,
Vettolani, G. 1998 \textit{A\&A} 334, 404 
\bibitem[\protect\citeauthoryear{Sylos Labini \& Montuori}{1998}]{SylosLabini} Sylos Labini, F., Montuori, M., 1998, Astron. \& Astrophys., 331, 809
\bibitem[\protect\citeauthoryear{Sylos Labini, Montuori \& Pietronero}{1998}]{SylosLabini1} Sylos Labini, F., Montuori, M., Pietronero, L.,
1998, Phys. Lett., 293, 62
\bibitem[\protect\citeauthoryear{Sylos Labini, Vasilyev \& Baryshev}{2006}]{SylosLabini2} Sylos Labini, F., Vasilyev, N.L., Baryshev,
Y.V., Archiv.Astro.ph/0610938 
\bibitem[\protect\citeauthoryear{Tekhanovich \& Baryshev}{2016}]{Tekhanovich} Tekhanovich D.I.I and Baryshev Yu.V., Archiv.Astro.ph/1610.05206
\bibitem[\protect\citeauthoryear{Vettolani et al}{1993}]{Vettolani} Vettolani, G., et al., 1993, in: Proc. of Schloss Rindberg Workshop: Studying the Universe With Clusters of Galaxies
\bibitem[\protect\citeauthoryear{Wu, Lahav \& Rees}{1999}]{Wu} Wu, K.K.S., Lahav, O., Rees, M.J., 1999, Nature
397, 225
\end{thebibliography}
\end{document}